\documentclass[conference]{IEEEtran}
\IEEEoverridecommandlockouts
\usepackage{cite}
\usepackage{amsmath,amssymb,amsfonts,amsthm}
\usepackage{graphicx}
\graphicspath{{./figures/}}
\usepackage{subfigure}
\usepackage{textcomp}
\usepackage{mathtools}
\usepackage{mathptmx}
\DeclarePairedDelimiter\ceil{\lceil}{\rceil}

\usepackage[utf8]{inputenc}
\usepackage[english]{babel}
\usepackage{url}

\usepackage{xcolor}
\usepackage{algorithm,algcompatible,amsmath}
\algnewcommand\INPUT{\item[\textbf{Input:}]}
\algnewcommand\OUTPUT{\item[\textbf{Output:}]}

\def\BibTeX{{\rm B\kern-.05em{\sc i\kern-.025em b}\kern-.08em
    T\kern-.1667em\lower.7ex\hbox{E}\kern-.125emX}}
\begin{document}

\title{Communication Contention Aware Scheduling of Multiple Deep Learning Training Jobs}

\author{\IEEEauthorblockN{Qiang Wang, Shaohuai Shi, Canhui Wang, Xiaowen Chu\IEEEauthorrefmark{1}\thanks{*Corresponding author.}}
\IEEEauthorblockA{Department of Computer Science, Hong Kong Baptist University
       \\ \{qiangwang, csshshi, chwang, chxw\}@comp.hkbu.edu.hk
       }
}

\maketitle

\begin{abstract}
Distributed Deep Learning (DDL) has rapidly grown its popularity since it helps boost the training performance on high-performance GPU clusters. Efficient job scheduling is indispensable to maximize the overall performance of the cluster when training multiple jobs simultaneously. However, existing schedulers do not consider the communication contention of multiple communication tasks from different distributed training jobs, which could deteriorate the system performance and prolong the job completion time. In this paper, we first establish a new DDL job scheduling framework which organizes DDL jobs as Directed Acyclic Graphs (DAGs) and considers communication contention between nodes. We then propose an efficient algorithm, LWF-$\kappa$, to balance the GPU utilization and consolidate the allocated GPUs for each job. When scheduling those communication tasks, we observe that neither avoiding all the contention nor blindly accepting them is optimal to minimize the job completion time. We thus propose a provable algorithm, AdaDUAL, to efficiently schedule those communication tasks. Based on AdaDUAL, we finally propose Ada-SRSF for the DDL job scheduling problem. Simulations on a 64-GPU cluster connected with 10 Gbps Ethernet show that LWF-$\kappa$ achieves up to $1.59\times$ improvement over the classical first-fit algorithms. More importantly, Ada-SRSF reduces the average job completion time by $20.1\%$ and $36.7\%$, as compared to the SRSF(1) scheme (avoiding all the contention) and the SRSF(2) scheme (blindly accepting all of two-way communication contention) respectively.
\end{abstract}


\section{Introduction}
The popularity of Deep Neural Network (DNN)\cite{lecun2015deep} grows rapidly in both industry and academia with its significant role in various applications, such as image classification, object detection, speech recognition, etc. 
As the training data size gets larger, it is natural to distribute the training workload to clusters and exploit the parallel computing power to increase the system's throughput. Distributed Deep Learning (DDL)\cite{lsddn2012}, especially with the data-parallel synchronous stochastic gradient descent (S-SGD) method, is typically adopted to speed up the training procedure. The success of DDL based on the data-parallel S-SGD method comes from using a large training batch size\cite{goyal2017accurate,jia2018highly} by leveraging the great computing power of hardware accelerators in a cluster, such as GPUs and TPUs. Nevertheless, compared to the fast-evolving computing capability of hardware accelerators and the intra-node interconnect technique like NVLink \cite{nvlink}, the inter-node communication of clusters has a much slower growth and gradually becomes the system bottleneck\cite{poseidon2017,shi2018performance}. Furthermore, if the cluster resources are shared among multiple DDL training jobs, network contention will frequently happen and prolong the communication time significantly, which consequently degrades the system performance in terms of training speed and resource utilization. 

Some recent studies have developed efficient DDL training algorithms to reduce the communication overhead by pipelining between computation and communication \cite{mgwfbp2019,blueconnect2019,jia2018highly,shi2020communication} and model/gradient compression \cite{lin2017deep,shi2019distributed}. However, when multiple DDL training jobs are running on a cluster, the network contention could lead to severe performance degradation if those communication tasks are not scheduled properly. For such a system that concurrently handles a rising number of jobs, flexible resource allocation and efficient job scheduling are indispensable to maximize the resource utilization.  

There exist many DDL training frameworks \cite{litz2018, lowPrep2019,tony2019,optimus2018,tiresias2019,gandiva2018} for CPU or GPU clusters, which allow users to share the resources and run their jobs concurrently. Some traditional schedulers \cite{plan2015,network2019,tetri2016,makespan2019} are not specifically designed for DDL training jobs and cannot leverage the characteristics of DDL (such as iterativeness and convergence properties) for maximal training efficiency. Some recent studies\cite{cynthia2019,bao2018,liu2019} focus on resource provisioning for multiple DDL jobs on the cloud. Besides, energy efficient scheduling algorithms for GPU clusters \cite{xin2017scheduling,chau2017scheduling} also raises attention since heavy deep learning workload on GPU clusters can bring tremendous energy consumption. However, we find that most of the schedulers do not consider the communication contention during the runtime job execution. To understand the impact of communication contention on the training performance, we have conducted an empirical study on a small cluster of four-GPU nodes connected by 10 Gbps Ethernet. When we execute only one DDL job with four GPUs on the cluster, the job completion time is 295 seconds. However, when we concurrently execute four same DDL jobs, each of which still uses four GPUs but from different nodes, the job completion time dramatically increases to 675 seconds due to the extensive communication contention. 

In this paper, we propose a new DDL job scheduling framework that explicitly considers the communication overhead and the communication contention among nodes. To address the communication contention issue, we develop a two-stage solution to the problem, including job placement and job scheduling. 
%The most interesting thing is that neither avoiding all the communication contention nor accepting all of them is optimal, which needs an adaptive algorithm to schedule those communication tasks. 
Our contributions are summarized as follows:
\begin{itemize}
    \item We establish a new DDL job scheduling framework which organizes the DDL jobs as DAGs. The framework assumes that each node/server has limited network resources and takes into account the communication contention when there are more than one tasks competing for network resources. To the best of our knowledge, this is the first work that considers the contention of multiple communication tasks launched by different DDL jobs. 
    \item We propose new efficient algorithms for the proposed framework in two stages, namely job placement and job scheduling. For job placement, we derive an LWF-$\kappa$ scheme to balance the intra-node and inter-node overhead of computation and communication. For job scheduling, we develop a provable algorithm, AdaDUAL, to adaptively schedule the communication tasks with potential contention. Based on AdaDUAL, we propose Ada-SRSF to schedule online multiple DDL training jobs.
    \item Supported by the measured data on real hardware, we conduct extensive simulation studies on a cluster of 16 four-GPU servers (i.e., 64 GPUs in total), connected by a 10Gbps Ethernet switch, to evaluate the performance of LWF-$\kappa$ and Ada-SRSF. The experimental results show decent improvement in terms of average job completion time and GPU resource utilization. We implement the communication contention script and a prototype of our DDL job scheduling framework on PyTorch, and make them publicly available\footnote{All experimental settings and source codes can be found at GitHub: \\ \url{https://github.com/HKBU-HPML/dl-scheduling}}.
\end{itemize}

The rest of the paper is organized as follows. We present some preliminaries in Section \ref{sec:preliminaries}, followed by the formulation of the DDL job scheduling problem in Section \ref{sec:problemformulation}. We then divide the problem into two stages, placement and scheduling, and propose communication contention aware algorithms, LWF-$\kappa$ and Ada-SRSF respectively for those two stages in Section \ref{sec:solution}. After that, we demonstrate our simulation setup as well as the performance evaluation for the proposed algorithms in Section \ref{sec:experiments}. Section \ref{sec:relatedwork} presents the related work, and Section \ref{sec:conclusion} concludes the paper and discusses the future work.

\section{Preliminaries}\label{sec:preliminaries}
\subsection{Distributed Deep Learning Training}\label{bg:ddl}
S-SGD: The DNN model is trained in an iterative manner with the target of minimizing a loss function $\mathcal{L}(W, D)$, where $W$ and $D$ are respectively the model weights and the input data. For large-scale DNNs, the data-parallel synchronized SGD (S-SGD) is widely applied to train models with multiple workers (say $N$ workers, and indexed by $g$) because it has the same convergence performance as the sequential SGD. Generally the ${i^{th}}$ iteration of the training contains four steps: a)  Each worker $g$ loads a mini-batch of local data $D_{i}^g$ into the device memory. b) Each worker $g$ performs a feed forward on $D_{i}^g$ through the neural network and computes the value of the loss function $\mathcal{L}(W_i, D_{i}^g)$. c) The first order gradients w.r.t. $W_i$ are calculated by backpropagation. d) Gradients from all the workers $\nabla\mathcal{L}(W_i, D_{i}^g)$ are aggregated, averaged and then distributed, which is often tackled by the All-Reduce collective function. Then all the workers update the model as Eq. \eqref{eq:ssgd_update}. 
\begin{align}
    W_{i+1} = W_{i}-\xi \frac{1}{N}\sum_{g=1}^{N}\nabla\mathcal{L}(W_i, D_{i}^g) \label{eq:ssgd_update}
\end{align}
The most common scenario of DDL training is using a large number of computing devices distributed among nodes in a cluster. As a result, the step d) involves extra communication overheads, which may become the bottleneck of DDL training and lower the resource utilization. 

%\subsection{Multiple DL Job Scheduling}
%Based on characteristics of the job: how many GPU(spatial), how long(temporal). Scheduling in two dimensions: where(spatial), when(temporal)
\subsection{Communication Model}
In Eq. \eqref{eq:ssgd_update}, we use $\Delta W_i = \frac{1}{N}\sum_{g=1}^{N}\nabla\mathcal{L}(W_i, D_{i}^g)$ to represent the aggregation of gradients from $N$ workers, which is an all-reduce operation. There are many efficient algorithms for the All-Reduce operation with different number of processes and message sizes \cite{optimized_all_reduce,mpi_all_reduce,model_all_reduce}. For brevity, we assume that the number of nodes is power-of-two. The inter-node communication cost without contention can be modelled as $\alpha+\beta M$ \cite{alpha-beta}, where $\alpha$ is the latency component, $\beta$ is the transmission time per byte, and $M$ is the message size. Without loss of generality, we do not limit the communication model to one specific algorithm. Given a constant number of nodes $N$, the time cost of a single All-Reduce operation without contention can be generalized as
\begin{align}
	T_{ar}=a+bM \label{eq:t_comm} 
\end{align}
where $a$ and $b$ are two constant numbers that are not related to $M$. Some optimized All-Reduce algorithms are summarized in Table \ref{tab:all_reduce}.
\begin{table}[ht]
	\centering
	\caption{Cost of Different All-Reduce Algorithms}
	\centering
	\begin{tabular}{|l|c|c|} \hline
		All-Reduce Algorithm & $a$ & $b$ \\ \hline \hline
		Binary tree & $2\alpha \log N$ & $(2\beta+\gamma) \log N$ \\ \hline
		Recursive doubling & $\alpha \log N$ & $(\beta+\gamma) \log N$ \\ \hline
		Recursive halving and doubling & $2\alpha \log N$ & $2\beta-\frac{1}{N}(2\beta+\gamma)+\gamma$ \\ \hline
		Ring & $2(N - 1)\alpha$ & $\frac{2(N-1)}{N}\beta+\frac{(N-1)}{N}\gamma$ \\ \hline
	\end{tabular}
	\label{tab:all_reduce}
\end{table}
\section{System Modeling and Problem Formulation}\label{sec:problemformulation}
\subsection{System Modeling}
\subsubsection{GPU Performance Modeling}
As mentioned in Section \ref{bg:ddl}, during the DDL job training, the compute tasks using GPUs contain b) feed forward and c) backpropagation, which are the two main computation parts. We calculate their time consumption as follows.

Feed-forward time is denoted as $t_{f_k}$:
\begin{align}
    t_{f_k} &= \frac{{\lambda}_{f}B_k}{P},
    %\hat{P}_{i,j} &= P_{i,j}f_f(N_j)
\end{align}
and backpropagation time is denoted as $t_{b_k}$:
\begin{align}
	t_{b_k} &= \frac{{\lambda}_{b}B_k}{P},
	%\hat{P}_{i,j} &= P_{i,j}f_b(N_j)
\end{align}
where $\lambda_{f}$ and $\lambda_{b}$ are the workload coefficients relevant to the DNN model, $B_k$ is the mini-batch size, and $P$ is the theoretical peak performance of the GPU. In practice, we can measure $t_{f_k}$ and $t_{b_k}$ for a given model since they will not change during the entire training procedure. 

\subsubsection{Communication Contention Modeling}
\begin{figure}[!h]
	\centering
	\includegraphics[width=0.92\linewidth]{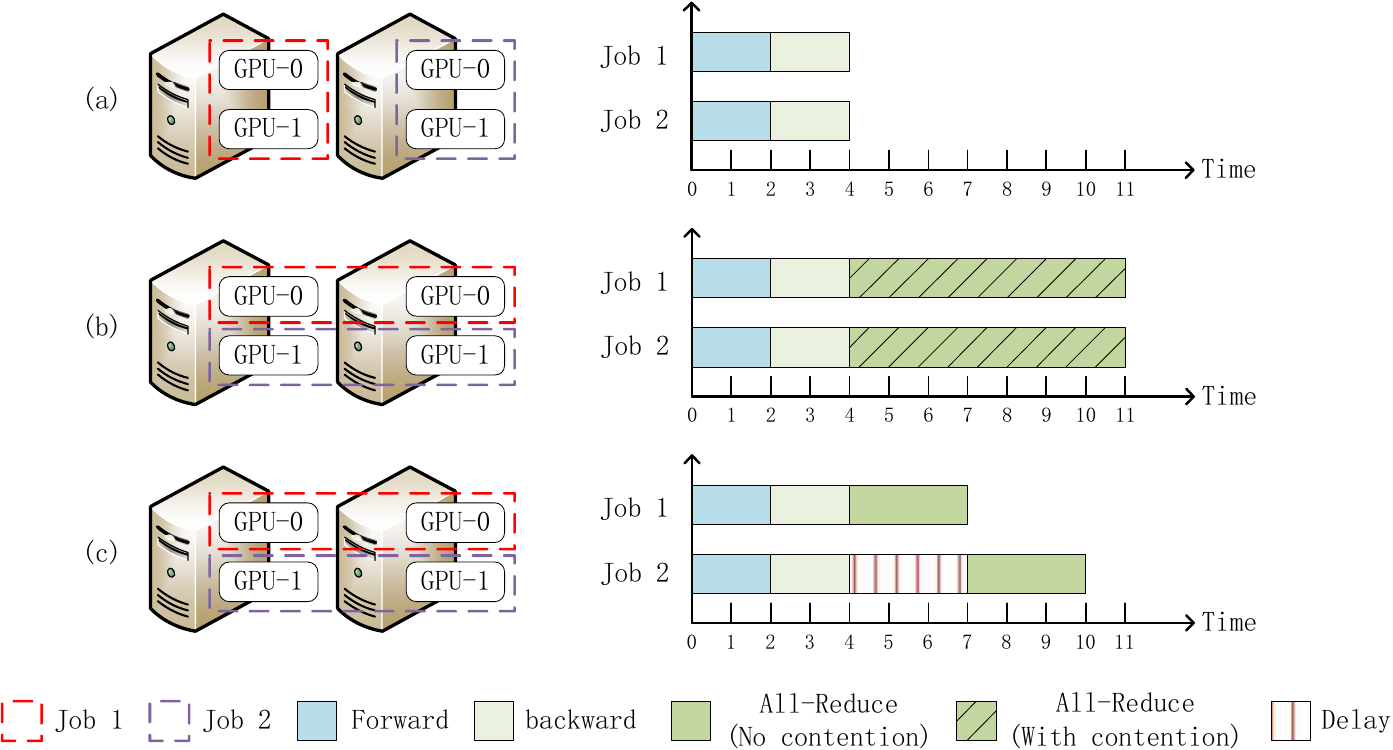}
	\caption{(a) No communication task. (b) Start the communication tasks at the same time. (c) Process the communication tasks in order.}
	\label{fig:net_confl}
\end{figure}

%The start-up cost is largely due to software overhead on the sending and the receiving nodes. The routing of messages between nodes is subsequently done in hardware using wormhole routing, which pipelines messages and incurs a very small extra overhead due to the distance between the order of the cost of an instruction. two nodes [17]. Typically, $\alpha$ is four to five orders of magnitude greater than $\beta_1$ where $\beta_1$ is on the order of the cost of an instruction.
Notice that the network resources of the servers are shared among different jobs. For each time slot, it is possible that a new communication task is scheduled to the node whose network resource is already occupied by other communication tasks, or an existing communication task exits. In this case, all the existing communication tasks may update the communication time and the transfer rate for the rest data. An example of the communication overhead of different scheduling solutions is demonstrated in Fig. \ref{fig:net_confl}. Solution (a) as shown in Fig. \ref{fig:net_confl}(a) allocates the GPUs in the same server for each job, which results in no communication overhead. This case usually happens when there are enough GPU resources. Solutions (b) and (c) allocate the GPUs across different servers for each job, which introduces communication tasks. These cases usually happen when there are not enough GPU resources (e.g., GPU memory) on a single server. In our real experiments, we find that the communication contention can cut down the practical network bandwidth and bring extra overheads. Thus, Solution (c) achieves shorter job completion time than (b) by postponing the communication task of Job 2 to avoid contention. 

\begin{figure}[!ht]
	\centering
	\subfigure[]
	{
	\includegraphics[width=0.44\linewidth]{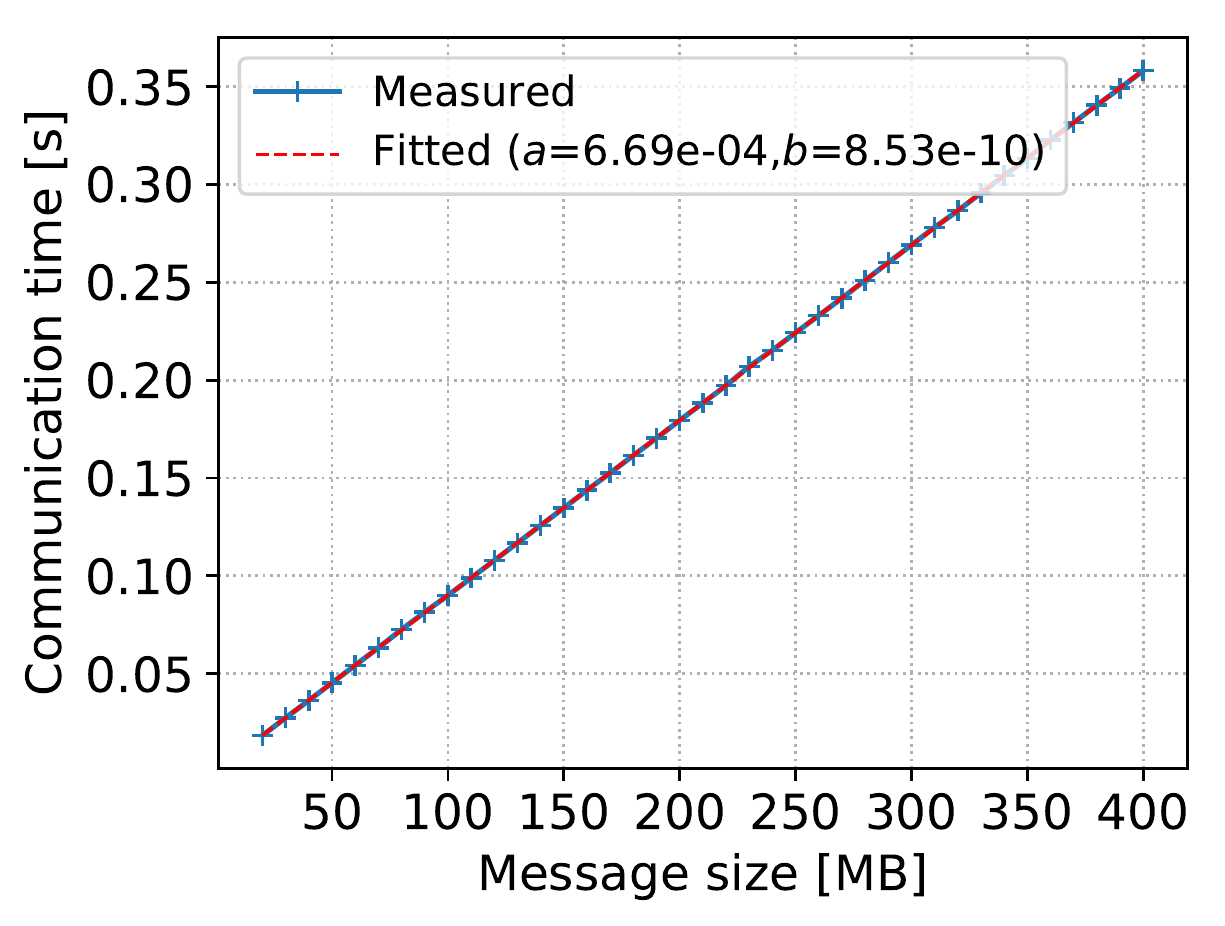}\label{fig:alpha-beta}
	}
	\subfigure[]
	{
	\includegraphics[width=0.43\linewidth]{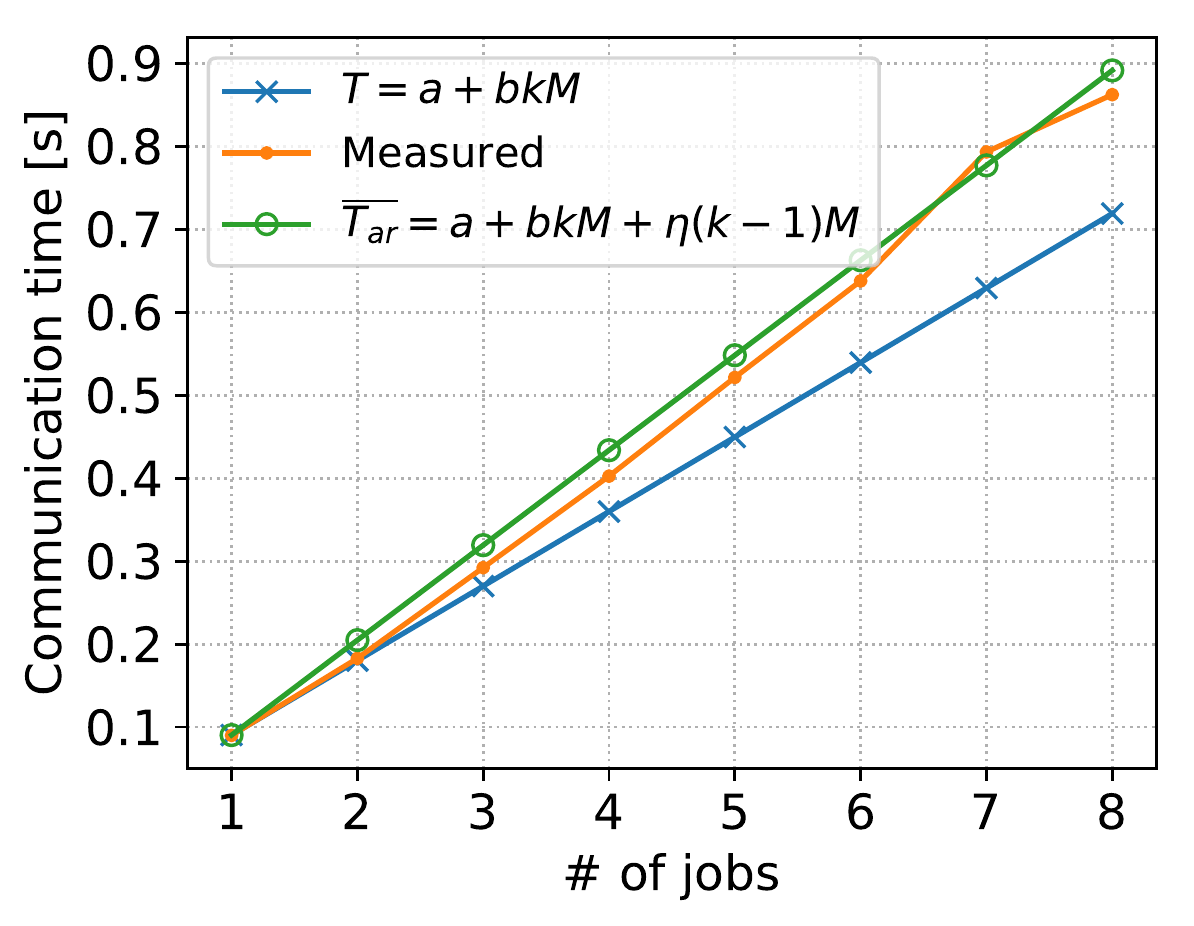}\label{fig:real_contention}
	}
	\caption{Communication performance models. (a) Single All-Reduce operation on two nodes. (b) Multiple All-Reduce operations (the message size is 100 MB) on two nodes.}
	\label{fig:comm_model}
	\vspace{-0.8 em}
\end{figure}

We conduct experiments of the ring-based All-Reduce algorithm on a 2-node cluster. The nodes are equipped with 10 Gbps Ethernet. We first measure only one communication task with different message size $M$ and fit Eq. \eqref{eq:t_comm}. The result is shown in Fig. \ref{fig:alpha-beta} and the estimated $a=6.69e-4$ and $b=8.53e-10$. Then we launch different number of communication tasks concurrently and measure the average time. The result is shown in Fig. \ref{fig:real_contention}. 
We model the time consumption of All-Reduce with contention as Eq. \eqref{eq:tc_comm}. The same $a$ and $b$ in Fig. \ref{fig:alpha-beta} is adopted. 
\begin{align}
	\overline{T}_{ar}&=a+kbM+(k-1){\eta}M, \label{eq:tc_comm}
\end{align}
where $k$ is the maximum number of concurrently running communication tasks over all the nodes. In the experiment of Fig. \ref{fig:comm_model}, we make all the communication tasks transfer the same message size between nodes and scale the number of jobs from 1 to 8, which corresponds to $k \in [1, 8]$ for all the nodes. Ideally, the network bandwidth is shared by the tasks in a round-bin manner, and the total time of transferring all the messages is $a+bkM$. However, we observe that the measured average communication time is higher than $a+bkM$ in Fig. \ref{fig:real_contention}. We hereby introduce an item $(k-1)\eta M$ to represent the extra overhead. When $k=1$, which means no contention, the $(k-1)\eta M$ vanishes and Eq. \ref{eq:tc_comm} reduces to Eq. \ref{eq:t_comm}.
%\begin{subequations}
%	\begin{align}
%	\overline{T}_{r}&=a+bkn+k{\eta}n \label{eq:tc_comm} \\
%	\overline{R}_{r}&= \frac{1}{a/n+bk+k\eta} \label{eq:rc_comm}
%	\end{align}
%\end{subequations}

Compared to Eq. \eqref{eq:t_comm}, the data transmission is enlarged by $k$ times and the total time consumption includes the penalty item $(k-1)\eta M$. $kbM$ indicates that the network bandwidth is shared evenly among the jobs. However, the network efficiency goes down due to the $(k-1)\eta M$ penalty. 
%Alternatively, links on which there is a conflict may be multiplexed, which yields the same modification to the cost. 

%Case 3: $\tau_{l,m}^{k}$ is gradient aggregation.
%\begin{align}
%e_{l,m}^k &= \text{max}\{\frac{M_k}{\hat{B}_{i}}+{\xi}N_j, \frac{M_k}{\hat{W}}+{\sigma}N_j\} \\
%\hat{W} &= W{f_W}(N_j), \hat{B}_i = B_i{f_B}(N_j)
%\end{align}

\subsection{Problem Formulation}
For ease of reference, we summarize some frequently used notations throughout this paper in Table \ref{tab:notation}.
\begin{table}[!ht]
	\centering
	\caption{Frequently used notations}
	\begin{tabular}{|p{0.3in}|p{2.7in}|} \hline
		Name & Descriptions \\ \hline \hline
		\textsl{$N_{s}$} & \# of servers of the cluster \\
		\textsl{$N_{g}$} & \# of GPUs on each server \\
		\textsl{$S_{i}$} & the $i$-th server of the cluster \\
		\textsl{$g_{i,j}$} & the $j$-th GPU on the $i$-th server \\
		\textsl{$P_{i,j}$} & the theoretical performance of $g_{i,j}$ (GFLOPS) \\
		%\textsl{$W_{i}$} & the practical bandwidth of $S_{i}$ \\
		%\textsl{$W$} & the practical bandwidth of the switch \\
		%\textsl{${P_{i,j}}$} & the theoretical peak performance of $g_{i,j}$ (GFLOPS) \\
		%		\textsl{${B_{i}}$} & the theoretical peak bandwidth of $S_{i}$ \\
		%		\textsl{${W}$} & the theoretical peak bandwidth of the switch \\
		\hline
		$\mathbb{J}$ & The job set \\
		$N_{\mathbb{J}}$ & \# of total jobs \\
		$J_{k}$ & the $k$-th job \\
		$\mathbb{G}(J_{k})$ & the set of GPUs used by $J_{k}$ \\
		$\mathbb{S}(J_{k})$ & the set of servers used by $J_{k}$ \\
		$A_{k}$ & the arrival time of $J_{k}$ \\
		$B_{k}$ & the mini-batch size used by $J_{k}$ \\
		$I_{k}$ & \# of iterations that $j_{k}$ needs to run \\
		$T$ & system timespan \\
		$E_{k}$ & the timestamp when $J_{k}$ is finished \\
		\hline
		$\mathbb{T}_{k}$ & the task set of $J_{k}$ \\
		$D_{l}^{k}$ & the $l$-th DAG in $\mathbb{T}_{k}$ \\
		$\tau_{l,m}^{k}$ & the $m$-th task of $D_{l}^{k}$ \\
		${\sigma}^{k}$ & the DNN model size of $J_{k}$ \\
		$f^{k}$ & the forward task of $J_{k}$ \\
		$b^{k}$ & the backward task of $J_{k}$ \\		
		$c^{k}$ & the communication task of $J_{k}$ \\
		%$u_{l,m}^{k}$ & the issue time stamp of $\tau_{l,m}^{k}$ \\
		%$e_{l,m}^{k}$ & the time used to execute task $\tau_{l,m}^{k}$ \\
		\hline
		$\lambda_{b},\lambda_{f}$ & the workload coefficients relevant to the DNN model \\
		$a, b, \eta$ & the parameters of the communication contention model \\
		\hline
	\end{tabular}
	\label{tab:notation}
% 	\vspace{-1.6 em}
\end{table}

\begin{figure}[!h]
	\centering
	\includegraphics[width=0.9\linewidth]{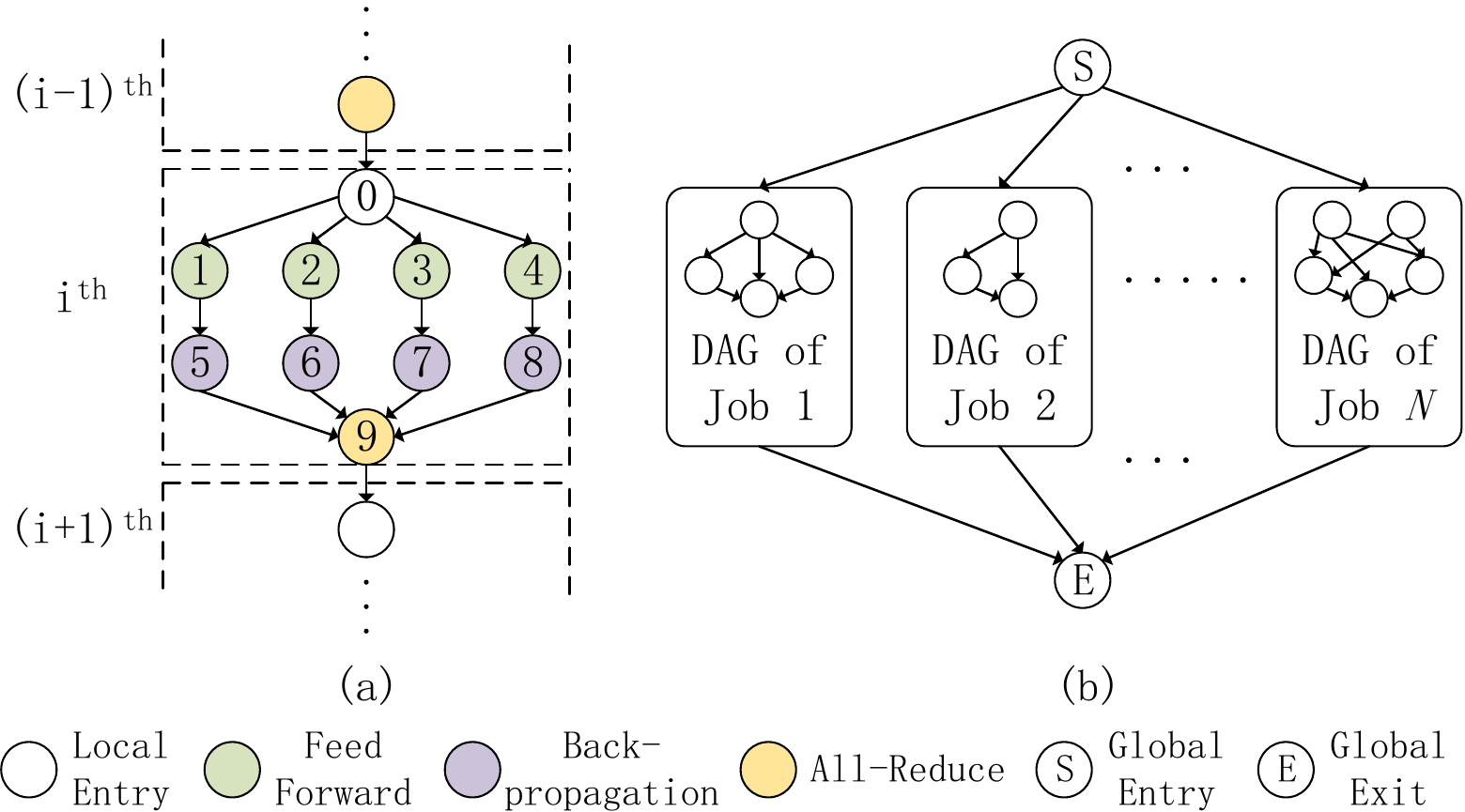}
	\caption{(a) The sub-DAG presentation of one DDL job. (b) The DAG presentation of multiple DLL jobs. The global entry is followed by all the local entries of the first iteration, while the global exit is the successor of all the All-Reduce task of the final iteration.}
	\label{fig:dag}
\end{figure}

We first establish a new DLL job scheduling framework which organizes each job as a directed acyclic graph (DAG). An example of a DDL job using four workers (e.g., GPUs) is illustrated in Fig. \ref{fig:dag}(a). One training iteration in a job can be represented by a child DAG. The child DAG includes three types of tasks, feed forward, backpropagation, and All-Reduce. The feed forward and backpropagation of different workers can be executed independently, while the All-Reduce task has a synchronization barrier to wait for all the backpropagation tasks to finish. In each child DAG, the feed forward task is the entry task and the gradient aggregation is the exit task. If the job runs for $R_{k}$ iterations to achieve a convergent model, the entire DAG is composed of $R_{k}$ identical child DAGs, in which the All-Reduce task of the $i$-th iteration follows the entry task of the $(i+1)$-th iteration. 

Our DDL job scheduling problem arises from the following system setting and assumptions:
\subsubsection{The GPU cluster}
A GPU cluster consists of $N_s$ servers, $S_i, i \in [1, N_s]$, connected with a network switch of sufficiently large bandwidth. Each server $S_i$ has $N_g$ GPUs. $g_{i,j}$ denote the $j$-th GPU on the $i$-th server. Assume that all the GPUs have the same specification and theoretical peak performance, denoted by $P_{i,j}$. The network performance of each server is modelled by Eq. \eqref{eq:tc_comm} and shared among different jobs. 

\subsubsection{DDL job characteristics}
A job set $\mathbb{J}$ of $N_{\mathbb{J}}$ training jobs arriving over time. The job $J_k$ arrives at $A_k$. Each job can be represented by a DAG. We assume that the tasks in each job are non-preemptive, and the job is preemptive if the ongoing task is finished.
%A DL training job contain $R_k$ iterations, each of which tackles the same tasks on a mini-batch of data. One iteration of a job $J_k$ contains 6 tasks: (a) Read a mini-batch of data from disk to memory. (b) Transfer the data from the CPU memory to the GPU memory. (c) GPU kernels are launched to do feed forward operation layer by layer. (d) To minimize the objective function, first order gradients w.r.t weights and inputs are calculated with chain rule, which is the phase of backward propagation. (e) Gradients from all GPUs are aggregated and averaged in one node, which is the phase of gradient aggregation, and it can be done by CPU or GPU. (f) The model is updated based on the aggragated gradients. Here we re-arrange the procedure of (a)$\sim$(f) by moving (f) in front of (c) and merge (f) into (c), since weight update is usually much faster than the feed forward. For the first iteration, we regard the weight initialization as the weight update task. Here we only discuss (c)$\sim$(e), since the data loading in one node is usually fast and can be overlapped by the feed forward and backward propagation. 

\subsubsection{The allocated GPUs for each job}
$\mathbb{G}(J_k)$ denotes the set of GPUs used by $J_k$ and it can be within the same server or across different servers. $\mathbb{G}(J_k)$ will not change for each job. Each GPU can only be occupied by one job at any time slot.

%The objective function is described in Eq. \eqref{eq:obj_func}, which means that we want to minimize the total execution time of all the jobs.  
%\begin{align}\label{eq:obj_func}
%\underset{k \in [1,2,...,I]}{\text{minimize}} \sum{E_{k}}
%\end{align}

Given a job set of DDL training, we first build a DAG for each of them and integrate those small DAGs into a large one, as illustrated in Fig. \ref{fig:dag}(b). We introduce a virtual global exit task and all the All-Reduce tasks of the last iteration connect to it. The global DAG is denoted by $G = (V, E)$, where $V$ is the set of $v$ tasks and $E$ is the set of $e$ edges between tasks. Each task has its time consumption and each edge $i, j\in E$ represents the procedure constraint such that task $n_i$ should complete its execution before task $n_j$ starts. We assume that the edge has no extra overhead since we regard the communication between nodes as a concrete task in $G$. 

Our objective is to minimize the average job completion time (JCT) of all the DDL jobs, i.e.,
\begin{align}
    \text{min.}& \sum_{k} (F_k - A_k), k \in [1, N_{\mathbb{J}}], \label{eq:avg_JCT}
\end{align}
where $F_k$ is the job completion time stamp of $J_k$.

\section{Solution}\label{sec:solution}
In this section, we describe our solution to solve the problem formulated in Section \ref{sec:problemformulation}. The solution comprises of two stages, job placement and job scheduling. Most of the traditional job scheduling solutions first decide the job scheduling order (job priority) and then assign them to some machines, while ours is different. Since one job may use more than one nodes, which requires network communication, the time cost of the All-Reduce operation cannot be determined before we know the selected GPUs and nodes of the job. Similarly, the possible network contention cannot be predicted before we allocate the GPUs to the jobs. For job placement, we develop an efficient algorithm, LWF-$\kappa$, which balances the intra-node and inter-node workload of computation and communication. For job scheduling, we first give the optimal solution with theoretical guarantees to the problem of scheduling two new-coming communication tasks. Then we derive an efficient algorithm, AdaDUAL, to schedule those communication tasks, and integrate it into our final communication contention aware job scheduling algorithm, which is called Ada-SRSF.

\subsection{Placement}
We assume that all the tasks of a DDL job are assigned to a fixed set of processors (CPUs/GPUs) once it has been scheduled since the frequent job preemption and migration may lead to severe performance degradation \cite{tiresias2019}. Before we design the scheduling algorithm, each new-arriving DDL job should be placed to a certain set of intra-node or inter-node processors, which is called job placement. 

However, there are two concerns when designing the job placement algorithm for our problem. First, if a job is placed among a large number of nodes, although it can achieve better utilization of those scattered processors, large communication overhead may bring worse system performance. Second, always addressing the consolidation of processors for placing a job may cause workload unbalance and processor time fragmentation among nodes. Thus, we design a fine-grained job placement algorithm, LWF-$\kappa$, which makes trade-off between resource utilization and communication avoidance. LWF-$\kappa$ first decides the job priority according to the job training time and its used GPU number. Then LWF-$\kappa$ picks up a set of GPUs from the cluster and assigns the job to them. 

\textbf{GPU Allocation.} Given a job $J_k$ that needs $n$ GPUs, if there are enough GPU resources in the cluster, such as GPU memory size, we must determine how to allocate them. The classical heuristic algorithms include First-Fit (FF) \cite{scheduling2011} and List-Scheduling (LS) \cite{scheduling2011}. FF chooses the first $n$ available GPUs in order, while LS chooses the top-$n$ GPUs with the least workload. However, those chosen GPUs by FF and LS are likely distributed among a large number of nodes, and the communication overhead may finally cause severe performance degradation. Thus, we propose a new placement algorithm, called LWF-$\kappa$, to balance resource utilization and communication avoidance. LWF indicates \textbf{l}east \textbf{w}orkload \textbf{f}irst, which is similar to LS. The difference between LWF-$\kappa$ and the classical LS is that $\kappa$ can be used to consolidate the GPUs for a job to some extend. Specifically, if $n <= \kappa$, we choose the top-$n$ GPUs for the job, which is the same as LS. In this way, those chosen GPUs is distributed among at most $\kappa$ nodes with controllable communication overhead. If $n > \kappa$, we first sort the servers by their total GPU workloads and then pick up those available GPUs server by server. 
%Obviously a bigger $\kappa$ results in a more aggressive mode to search those GPUs with the least workload. 

To explain the whole placement algorithm, we hereby define some terms used in our job placement solution. For each job $J_{k}$, we define $C_{J_{k}}$ as the total time consumption of executing the computation workload of DDL training, which is calculated by Eq. \eqref{eq:comp_job}. The computation workloads include forward feed and backpropagation which repeat for $I_{k}$ iterations. For the network connections between servers, we define $E_{J_k}$ as the total time consumption of executing all the communication tasks without contention, which is calculated by Eq. \eqref{eq:comm_server}. 
\begin{align}
    C_{J_k} &= (t_{f^k}+t_{b^k})\times I_{k}, \label{eq:comp_job} \\
    E_{J_k} &= 
    \begin{cases}
    0, & \text{if $|\mathbb{S}_{J_k}|=1$}  \\
    (a + b\sigma_{k}) \times I_{k},              & \text{otherwise}. 
    \end{cases} 
    \label{eq:comm_server}
\end{align}
We denote the remaining workload of each GPU as $L_{g_{i, j}}$, and the remaining workload of each job $J_k$ as $L_{J_k}$. We initialize $L_{g_{i, j}}=0$ and $L_{J_k}=(C_{J_k}+E_{J_k}) \times |\mathbb{G}(J_k)|$.

\textbf{Job Priority.} It is possible that several new DLL jobs are arriving at the same time (or within a shorter time interval than that of the minimum system time unit). The priority to allocate GPU resources to them should be also decided. We simply apply those size-based, length-based or both heuristic strategies in \cite{tiresias2019}, since they have been studied intensively. The size-based heuristic strategies determine the job priority according to their used GPU numbers, while the length-based ones determine the job priority according the job running time. We apply the shortest-remaining-service-first (SRSF) algorithm since it performs well most of time. Since we cannot decide the communication workload of one job before it is allocated on some GPUs, we set $E_{J_k}=0$ when sorting the jobs by SRSF.

We present our efficient placement algorithm LWF-$\kappa$ in Algorithm 1. Line 1 conducts some initialization of the job workload. Lines 2-9 determine the GPU set for the given job when $|\mathbb{G}_J| \le \kappa$. The solution is just the same as LS, which picks up the top-$\mathbb{G}_{J}$ with the least workload. Lines 10-21 determine the GPU set for the given job when $|\mathbb{G}_J| > \kappa$. Lines 11-15 sort the servers by their total remaining workload and pick up the available GPUs from them one by one. Lines 6 and 18 update the GPU remaining workload if they are successfully allocated to the job. Line 22 returns an empty set if the available GPU set cannot be found due to the GPU memory limitation. 
\begin{algorithm}
    \caption{LWF-$\kappa$ Job Placement}
  \begin{algorithmic}[1]
    \INPUT A given job $J$ to allocate GPUs, the current remaining workload of all the GPUs $L_{g_{i, j}}$, the current total remaining workload of each server $L_{S_i}=\sum_{j}L_{g_{i, j}}$.
    \OUTPUT The GPU set of the job $J$, $\mathbb{G}(J)$
    \STATE \textbf{Initialize} $L_J=C_{J} \times |\mathbb{G}(J)|$.
    \IF {$|\mathbb{G}(J)| \le \kappa$}
    \STATE $\mathbb{G}_{avail}$ $\leftarrow$ GPUs that have enough rest memory size.
    \IF {$|\mathbb{G}_{avail}| \ge |\mathbb{G}(J)|$}
    \STATE Select top-$|\mathbb{G}(J)|$ GPUs with the least $L_{g_{i, j}}$ from $\mathbb{G}_{avail}$ as $\mathbb{G}(J)$
    \STATE $L_{g_{i, j}}$+=$L_J$ for $g_{i, j} \in \mathbb{G}(J)$
    \STATE return $\mathbb{G}(J)$
    \ENDIF
    \ENDIF
    \IF {$|\mathbb{G}(J)| > \kappa$}
    \STATE Sort the $\mathbb{S}$ by their $L_{S_i}$, pick up the top-$\ceil*{|\mathbb{G}(J)|/N_g}$ servers as $\mathbb{S}_{chosen}$
    \STATE $\mathbb{G}_{avail} \leftarrow \emptyset$
    \FOR {$S_i$ in $\mathbb{S}$}
    \STATE Select those GPUs in $S_i$ that have enough rest memory size and sort them by $L_{g_{i, j}}$. Then append them to the tail of $\mathbb{G}_{avail}$
    \ENDFOR
    \IF {$|\mathbb{G}_{avail}| \ge |\mathbb{G}(J)|$}
    \STATE Select top-$|\mathbb{G}(J)|$ GPUs from $\mathbb{G}_{avail}$ as $\mathbb{G}(J)$
    \STATE $L_{g_{i, j}}$+=$L_J$ for $g_{i, j} \in \mathbb{G}(J)$
    \STATE return $\mathbb{G}(J)$
    \ENDIF
    \ENDIF
    \STATE return $\emptyset$
  \end{algorithmic}
\end{algorithm}

\textbf{Time complexity of LWF-$\kappa$.} The time consumption of LWF-$\kappa$ is dominated by the sorting operations when choosing the GPUs with least workload. If $|\mathbb{G}(J)| \le \kappa$, we need to choose top-$|\mathbb{G}_{J}|$ GPUs, which consumes $\theta$($N_s N_g {log}_2 (N_s N_g)$) for sorting, where $N_s N_g$ is the total number of GPUs in the cluster. Otherwise, we only need to sort $N_s$ servers and then $|\mathbb{G}_{J}|$ GPUs from top-$\ceil*{|\mathbb{G}(J)|/N_g}$ servers, which consumes much less time of $\theta$($N_s {log}_2 N_s$) (we neglect the sorting time of intra-server GPUs since $N_g \ll N_s$). In practice, the latter one happens more frequently since we generally do not set a big $\kappa$. Even sorting the servers firstly and then the GPUs of them often give satisfying results. In conclusion, LWF-$\kappa$ commonly has a time complexity of $\theta$($N_s {log}_2 N_s$).
%\begin{theorem}
%Text
%\end{theorem}
%\begin{proof}
%Text
%\end{proof}

\subsection{Scheduling}
For those jobs which have been allocated GPU resources, we schedule each task in the DAG of each job at the specific time slot with the target of minimizing the average job completion time. Since we break an entire DDL job to a set of tasks grouped by a DAG, we have better flexibility to schedule them and take advantage of overlapping the computation and communication of different jobs. As for those ready forward and backward tasks, one GPU can select one of them by some heuristic rules and execute it. However, as for the communication tasks, since the network resources are shared among the jobs, it is possible to tackle multiple communication tasks at the same time slot. To simplify the solution derivation, let us first consider a problem of two communication tasks arriving at the same time slot. 

\textbf{Problem 1 (P1)} Assume that two communication tasks $c_1$ and $c_2$ arrive at time $T$ and their message sizes are $M_1$ and $M_2$ ($M_1 <= M_2$). The communication overhead is modelled by Eq. \eqref{eq:tc_comm}. We assume that $M_1$ and $M_2$ are sufficiently large and neglect the parameter $a$ in Eq. \eqref{eq:tc_comm}. If no contention happens, the communication time of $c_1$ and $c_2$ is respectively $t_1=bM_1$ and $t_2=bM_2$. For ease of calculation, we set $T=0$. Two cases to execute $c_1$ and $c_2$ are:

\textbf{Case 1 (C1):} $c_1$ at $t=0$, and then $c_2$ at time $t$, $t \in [0, t_1]$.

\textbf{Case 2 (C2):} $c_2$ at $t=0$, and then $c_1$ at time $t$, $t \in [0, t_2]$.

Assume that their ending time is $T_1$ and $T_2$. The target is to minimize their average completion time as Eq. \eqref{eq:comm_schedule}.
\begin{align}
    t_{aver}=(T_1+T_2)/2. \label{eq:comm_schedule}
\end{align}

\textbf{Theorem 1} Considering \textbf{C1} under \textbf{P1}, $t=t_1$ gives the minimum $t_{aver}$. 
\begin{proof}
We calculate $t_{aver}$ for \textbf{C1} as Eq. \eqref{eq:t_C1}. 
\begin{subequations}
\begin{align}
    T_1 &= t + 2b(M_1 - \frac{t}{b}) + \eta(M_1 - \frac{t}{b}) \label{eq:t_C1_T1} \\
    T_2 &= t + 2b(M_1 - \frac{t}{b}) + \eta(M_1 - \frac{t}{b}) + b(M_2 - (M_1 - \frac{t}{b})) \label{eq:t_C1_T2} \\
    t_{aver} &= (2t + 4b(M_1 - \frac{t}{b}) + 2\eta(M_1 - \frac{t}{b}) + b(M_2 - (M_1 - \frac{t}{b}))) / 2 \nonumber \\
                  &= (-(1+\frac{2\eta}{b})t+(3b+2\eta)M_1 + bM_2) / 2 \label{eq:t_C1} 
\end{align}
\end{subequations}
It is observed that $t_{aver}$ is monotonically decreasing with respect to $t$, which concludes that $t=t_1$ gives the minimum $t_{aver}$, which we denote by $\hat{t}_{aver}^{C1}$. Since $t_1 = bM_1$, $\hat{t}_{aver}^{C1}=(2bM_1 + bM_2)/2$.
\end{proof}
\textbf{Theorem 2} Considering \textbf{C2} under \textbf{P1}. If $\frac{M1}{M2}<\frac{b}{2(b+\eta)}$, \textbf{C2} with $t=0$ gives the minimum $t_{aver}$. Otherwise, \textbf{C2} with $t=t_2$ gives the minimum $t_{aver}$.
\begin{proof}
We divide the interval $[0, t_2]$ into two sub-intervals, $[0, b(M_2 - M_1)]$ and $(b(M_2 - M_1), t_2]$. $t \in [0, b(M_2 - M_1)]$ refers to the case that the whole message of $c_1$ is under contention, while $t \in (b(M_2 - M_1), t_2]$ refers to the case that either part or none of the message of $c_1$ is under contention. Then we derive $t_{aver}$ for them case by case.

(a) For $t \in [0, b(M_2 - M_1)]$, we have:
\begin{subequations}
\begin{align}
    T_1 &= t + 2bM_1 + \eta M_1 \label{eq:t_C2_1_T1} \\
    T_2 &= t + 2bM_1 + \eta M_1 + b(M_2 - M_1 - \frac{t}{b}) \label{eq:t_C2_1_T2} \\
    t_{aver} &= (2t + 4bM_1 + 2\eta M_1 + b(M_2 - M_1 - \frac{t}{b})) / 2 \nonumber \\
                  &= (t + (3b+2\eta)M_1 + bM_2) / 2 \label{eq:t_C2_1} 
\end{align}
\end{subequations}

It is observed that $t_{aver}$ is monotonically increasing with respect to $t$. In this case, $t=0$ gives the minimum $t_{aver}$, which we denote by $\hat{t}_{aver}^{C2a}$. $\hat{t}_{aver}^{C2a} = ((3b+2\eta)M_1 + bM_2)/2$ with $t=0$.

(b) For $t \in (b(M_2 - M_1), t_2]$, we have:
\begin{subequations}
\begin{align}
    T_1 &= t + 2b(M_2 - \frac{t}{b}) + \eta(M_2 - \frac{t}{b}) + b(M_1 - (M_2 - \frac{t}{b})) \label{eq:t_C2_2_T1} \\
    T_2 &= t + 2b(M_2 - \frac{t}{b}) + \eta(M_2 - \frac{t}{b}) \label{eq:t_C2_2_T2} \\
    t_{aver} &= (2t + 4b(M_2 - \frac{t}{b}) + 2\eta(M_2 - \frac{t}{b}) + b(M_1 - (M_2 - \frac{t}{b}))) / 2 \nonumber \\
                  &= (-(1+\frac{2\eta}{b})t+(3b+2\eta)M_2 + bM_1) / 2 \label{eq:t_C2_2} 
\end{align}
\end{subequations}

It is observed that $t_{aver}$ is monotonically decreasing with respect to $t$. In this case, $t=t_2$ gives the minimum $t_{aver}$, which we denote by $\hat{t}_{aver}^{C2b}$. Since $t_2 = bM_2$, $\hat{t}_{aver}^{C2b}=(bM_1 + 2bM_2)/2$.

Obviously, the minimum average completion time of $c_1$ and $c_2$ is either $\hat{t}_{aver}^{C2a}$ or $\hat{t}_{aver}^{C2b}$ under \textbf{C2}. Let $2(\hat{t}_{aver}^{C2a} -\hat{t}_{aver}^{C2b}) < 0$, we have
\begin{align}
    2(b+\eta)M_1 - bM_2 < 0 \nonumber \\
    \frac{M_1}{M_2} < \frac{b}{2(b+\eta)} 
\end{align}
which finally concludes that \textbf{C2} with $t=0$ gives smaller $t_{aver}$ than \textbf{C2} with $t=t_2$ if $\frac{M_1}{M_2}< \frac{b}{2(b+\eta)}$.
\end{proof}

Combining \textbf{Theorem 1} and \textbf{Theorem 2}, we compare three candidate minimums, $\hat{t}_{aver}^{C1}$, $\hat{t}_{aver}^{C2a}$ and $\hat{t}_{aver}^{C2b}$, as follows:
\begin{subequations}
\begin{align}
    \hat{t}_{aver}^{C1} &= (2bM_1 + bM_2)/2 \\
    \hat{t}_{aver}^{C2a}&= ((3b+2\eta)M_1 + bM_2)/2 \\
    \hat{t}_{aver}^{C2b}&= (bM_1 + 2bM_2)/2
\end{align}
\end{subequations}
It is obvious that $\hat{t}_{aver}^{C1}$ is smaller than the other two ($2b < (3b+2\eta)$, $b(M_1 - M_2) <= 0$). If there are two new arriving communication tasks, or the message size of the new arriving task is larger than the rest of the existing one, the best solution is to finish the smaller one first and then start the larger one. Besides, if the message size of the new arriving task is smaller than the rest of the existing one, \textbf{Theorem 2} is used to judge whether to start the new task immediately or wait until the existing one is finished. 
%In the practical production, it is more common that those communication tasks using the same links are not arriving at the same time slot. Theorem 1 can be also adapted to this case by calculating the transferred message size of the old tasks and updating it with the remaining message size. 
To this end, we derive a novel algorithm, called AdaDUAL, for scheduling the new arriving communication tasks in an adaptive manner. 

Algorithm 2 shows the procedure of AdaDUAL for communication task selection. For a given new communication task, AdaDUAL first counts the maximum existing communication tasks on the servers used by the new one, as indicated from line 2 to 7. Then lines 8-10 just accept the new task if there is no old communication tasks. Lines 11-18 show the case of only one existing task and apply \textbf{Theorem 1} and \textbf{2} to decide whether to start the new communication task at the current time slot according to the message sizes of them. Lines 19-21 reject the new communication task if there are more than two existing tasks. The case of scheduling more than two communication tasks is more complicated, which we leave as the future work. However, in our experiments, we observe that the $k$-way ($k > 2$) communication contention can dramatically decrease the network bandwidth efficiency and significantly prolong the execution time of all the jobs. 

\textbf{Time complexity of AdaDUAL.} AdaDUAL is an efficient and effective algorithm to schedule the communication tasks. Since lines 4-5 and 8-21 in Algorithm 2 only have $\theta$(1) time complexity, the time consumption of AdaDUAL is dominated by scanning all the servers in the for-loop, which results in a time complexity of $\theta$($N_\mathbb{S}$). 
\begin{algorithm}
  \caption{AdaDUAL for communication task selection}
  \begin{algorithmic}[1]
    \INPUT The running communication task set $\mathbb{C}_{S_i}$ on each server $i$, $i \in [1, N_s]$, the new-coming communication task $c_{new}$, its used servers $\mathbb{S}$ and message size $M_{new}$, the network performance parameter $b$, $\eta$
    \OUTPUT Whether to start $c_{new}$ at the current time slot. 
    \STATE \textbf{Initialize} $max\_tasks$=0. $\mathbb{C}_{old}=\emptyset$
    \FOR{$S$ in $\mathbb{S}$}
        \IF{$|\mathbb{C}_S| > max\_task$}
              \STATE $max\_task=|\mathbb{C}_S|$
              \STATE $\mathbb{C}_{old}=\mathbb{C}_{old}\cup \mathbb{C}_S$
        \ENDIF
    \ENDFOR
    \IF {$max\_task == 0$}
        \STATE return $True$ for starting $c_{new}$.
    \ENDIF
    \IF {$max\_task == 1$}
        \STATE Calculate the remaining message size $M_{old}$ of the current communication task in $\mathbb{C}_{old}$
        \IF{$\frac{M_{new}}{M_{old}}<\frac{b}{2(b+\eta)}$}
        \STATE return $True$ for starting $c_{new}$.
        \ELSE
        \STATE return $False$ for not starting $c_{new}$.
        \ENDIF
    \ENDIF
    \IF {$max\_task > 1$}
        \STATE return $False$ for not starting $c_{new}$.
    \ENDIF
%      \ENDFOR
%    \ENDWHILE
  \end{algorithmic}
\end{algorithm}

After deriving the solution to tackle communication contention, we combine AdaDUAL and the classical SRSF (shortest-remaining-service-first) \cite{tiresias2019} algorithm together to form our final communication contention aware job scheduling solution, called Ada-SRSF. Algorithm 3 shows the complete time-discrete procedure of Ada-SRSF. 
$Q$ is the set to store the queuing jobs. $R$ is the set to store the running jobs. The outer while loop in lines 2-31 terminates once all the jobs have been finished, which is equivalent to $U=\emptyset$. Line 3 updates the remaining workloads of all the GPUs and all the jobs. Line 4 checks whether there are some tasks or jobs that have been finished and removes them from $U$. Lines 6-13 allocate GPUs to those jobs in $Q$ using LWF-$\kappa$ if available. Lines 14-21 apply AdaDUAL to schedule those ready communication tasks. Lines 22-30 choose an available compute task, feed forward or backpropagation for each idle GPU. 
\begin{algorithm}
  \caption{The Ada-SRSF Communication Contention Aware Job Scheduling Algorithm}
  \begin{algorithmic}[1]
    \INPUT The job set $\mathbb{J}$, the arrival time of each job $A_k$, the GPU number used by each job $|\mathbb{G}(J_k)|$, $k \in [1, N_{\mathbb{J}}]$.
    %the job set that needs more than one nodes $\mathbb{J}_{comm}$
    \STATE \textbf{Initialize} $U=\mathbb{J}, t=0, Q=\emptyset, R=\emptyset, L_{g_{i, j}}=0, \newline L_{J_k}=(C_{J_k}+E_{J_k}) \times |\mathbb{G}(J_k)|$.
    \WHILE{$U \neq \emptyset$}
      \STATE {Update $L_{g_{i, j}}$ for all the GPUs in the cluster and $L{j_k}$ for all the jobs.}
      \STATE {Process the tasks leaving at the current time slot $t$. Remove them from $U$ and $R$.}
    \STATE {Append those tasks whose $A_k=t$ to the tail of $Q$. Sort $Q$ by the SRSF algorithm.}
    \STATE {\# Allocate GPUs to jobs in queue.}
    \FOR{$J$ in $Q$}
        \STATE Allocate GPUs for $J$ by LWF-$\kappa$
        \IF {$J$ is allocated successfully}
        \STATE {Update $L_{g_{i, j}}$ for those allocated GPUs.}
        \STATE {Append $J$ to the tail of $R$.}
        \ENDIF
    \ENDFOR
      \STATE {\# Schedule the communication tasks.}
      \STATE {Construct $\mathbb{J}_{comm}$, the set that contains DDL jobs ready for communication.}
      \STATE {Sort $\mathbb{J}_{comm}$ by the SRSF algorithm.}
      \FOR{$J$ in $\mathbb{J}_{comm}$}
        \IF {$J$ is not finished and ready for communication}
        \STATE Use AdaDUAL($J$) to decide whether to start All-Reduce of $J$ at $t$.
        \ENDIF
      \ENDFOR
      \STATE {\# Schedule compute tasks for all the GPUs.}
      \FOR{$i=1 \rightarrow N_s$}
        \FOR{$j=1 \rightarrow N_g$}
          \STATE  $\mathbb{A} \leftarrow$ the compute-ready job set on $g_{i, j}$
          \IF{$g_{i, j}$ is idle and $\mathbb{A} \neq \emptyset$}
              \STATE Sorting $\mathbb{A}$ by SRSF and select the available compute task $\tau$ from the first job, and then schedule it to $g_{i, j}$.
         \ENDIF
%          \ENDIF
        \ENDFOR
     \ENDFOR
    \ENDWHILE
  \end{algorithmic}
\end{algorithm}

\section{Performance Evaluation}\label{sec:experiments}
\subsection{Experimental Setup}
\textbf{Workload:} We generate the workload similar to the Microsoft job trace \cite{jeon2019analysis}. More details about the Microsoft trace can be found in \cite{jeon2019analysis} and Appendix of \cite{tiresias2019}. We generate totally 160 DDL jobs by scaling down the original job trace. As job characteristics, such as the number of GPUs and the training iterations, we mostly follow the distributions of the real trace: half of DDL jobs are single-GPU jobs; the rest are 14 2-GPU jobs, 26 4-GPU jobs, 30 8-GPU jobs, 8 16-GPU jobs, and 2 32-GPU jobs. The new distribution introduces more communication tasks than the original one. The training iteration of jobs varies from 1000 to 6000. We characterize our jobs based on both their number of GPUs and training iterations. We consider a job to be large if it needs more than 4 GPUs and otherwise small, and long if it runs for more than 1600 iterations and otherwise short. 
%Table \ref{tab:job_dist} lists the percentages of different job categories. 
%\begin{table}[!ht]
%	\centering
%	\caption{The Job Distribution by their number of GPUs (\textbf{S}mall and \textbf{L}arge) and their training iterations ($\mathbb{S}$hort and $\mathbb{L}$ong).}
%	\label{tab:job_dist}
%		\begin{tabular}{ccccc} \hline
%			Bin & 1 (\textbf{S}$\mathbb{S}$) & 2 (\textbf{S}$\mathbb{L}$) & 3 (\textbf{L}$\mathbb{S}$) & 4 (\textbf{L}$\mathbb{L}$) \\ \hline 
%			\% of Jobs & 11.8\%	& 63.1\% & 3.1\% & 21.9\% \\ \hline
%		\end{tabular}
%\end{table}

We simulate the task arrival of 160 jobs in 20 minutes and choose the basic time unit as one second, i.e. $T \in [1, 1200]$. We generate the number of arriving DDL jobs at each time slot, $n(T), T \in [1, 1200]$ according to the uniform distribution and refine it until $\sum_{T=1}^{1440}n(T)=160$. At each time slot, we pick the $\sum_{o=1}^{T-1}n(o)+1$-th to $\sum_{o=1}^{T}n(o)$-th jobs from the online job set to construct arrival jobs. 

\textbf{Cluster configurations:} We first conduct real experiments on a cluster of 16 servers to collect the data used for modeling communication contention. Each server has an Intel Xeon CPU E5-2670 and a 10Gbps
Ethernet (10GbE). The OS system is Ubuntu 16.04, and the MPI implementation is OpenMPI v3.1.0. For the GPU computational performance, we conduct real experiments on a single real Tesla V100-16GB GPU and collect the time consumption of feed forward and backpropagation of those tested DNNs. We run the DL job using PyTorch 1.2.0 \footnote{https://pytorch.org}.
%and Horovod 0.15.1\footnote{https://github.com/uber/horovod}. 
The DNN training parameters and performance on V100 are listed in Table \ref{tab:dnn_model}. 
\begin{table}[!ht]
	\centering
	\caption{DNN training parameter and performance on Tesla V100 GPU.}
	\label{tab:dnn_model}
		\begin{tabular}{|c|c|c|c|c|c|} \hline
			Network & Model & GPU Memory & Batch & $t_f(ms)$ & $t_b(ms)$\\ 
			& Size(MB) & Usage(MB) & Size & & \\ \hline\hline
			VGG-16 & 526.4	& 4527 & 16 & 35.8 & 53.7 \\ \hline
			ResNet-50 & 99.2 & 3213 & 16 & 25.0 & 37.4 \\ \hline
			Inception-V3  & 103.0 &	3291 & 16 & 34.9 & 52.4 \\ \hline
			LSTM-PTB  & 251.8 &	2751 & 64 & 31.5 & 47.3 \\ \hline
		\end{tabular}
\end{table}

\textbf{Simulation:} Based on the gathered experimental data on real hardware, we then simulate a cluster of 16 servers, each of which is equipped with four Tesla V100 GPUs. We implemented a discrete-time simulator to evaluate our placement and scheduling algorithms. 
%It simulates all job events in Tiresias, including job arrival, completion, demotion, promotion, and preemption.

\textbf{Baselines:} For job placement, we compare our LWF-$1$ algorithm with the random placement, denoted by RAND, and two classical heuristic algorithms, FF (First-Fit), and LS (List-Scheduling). We also explore the effects of different $\kappa$ when applying LWF-$\kappa$. For job scheduling, we compare our Ada-SRSF algorithm with SRSF($n$), where $n \in [1,2,3]$. SRSF(1) schedules those communication tasks under the constraint that each link between two nodes can be occupied by at most one task. SRSF(1) means no communication contention happening during the job execution. SRSF(2) allows each link between two nodes to be occupied by at most two tasks, which means a 2-way contention, without any constraint. Different from SRSF(2), Ada-SRSF allows a 2-way contention only if it helps reduce the job completion time. 

\textbf{Metrics:} For each experiments, we collect the JCT (job completion time) of all the DDL jobs as well as the utilization of all the GPUs. Then we plot the cumulative distribution curve of JCT. Besides, we also use several metrics, including average JCT, median JCT and 95th-percentile JCT and average GPU utilization, to compare different algorithms.

\subsection{Experimental Results}
\begin{figure*}[!ht]
	\centering
	\subfigure[]
	{
	\includegraphics[width=0.29\linewidth]{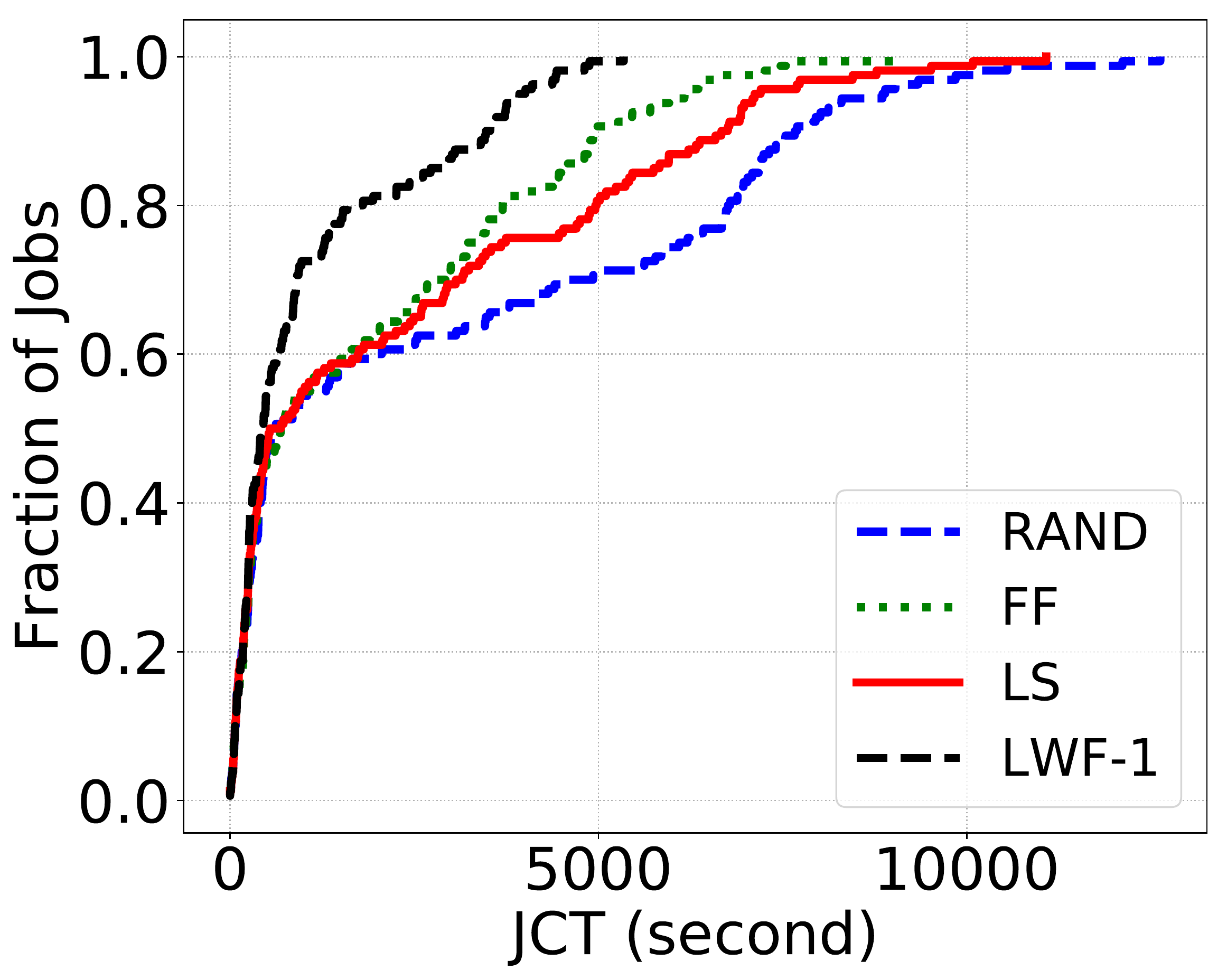}\label{fig:pg_a}
	}
	\subfigure[]
	{
	\includegraphics[width=0.285\linewidth]{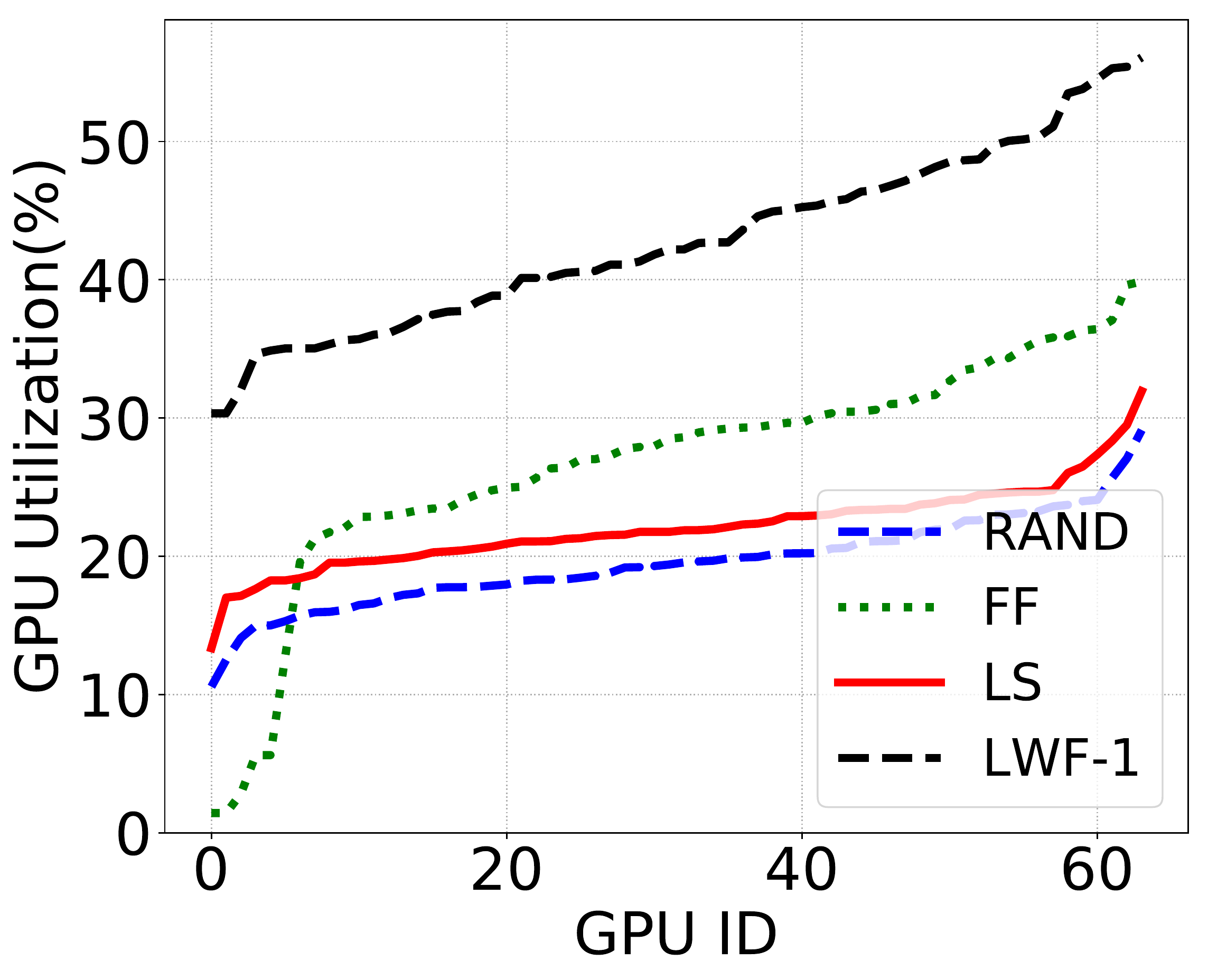}\label{fig:pg_b}
	}
	\subfigure[]
	{
	\includegraphics[width=0.30\linewidth]{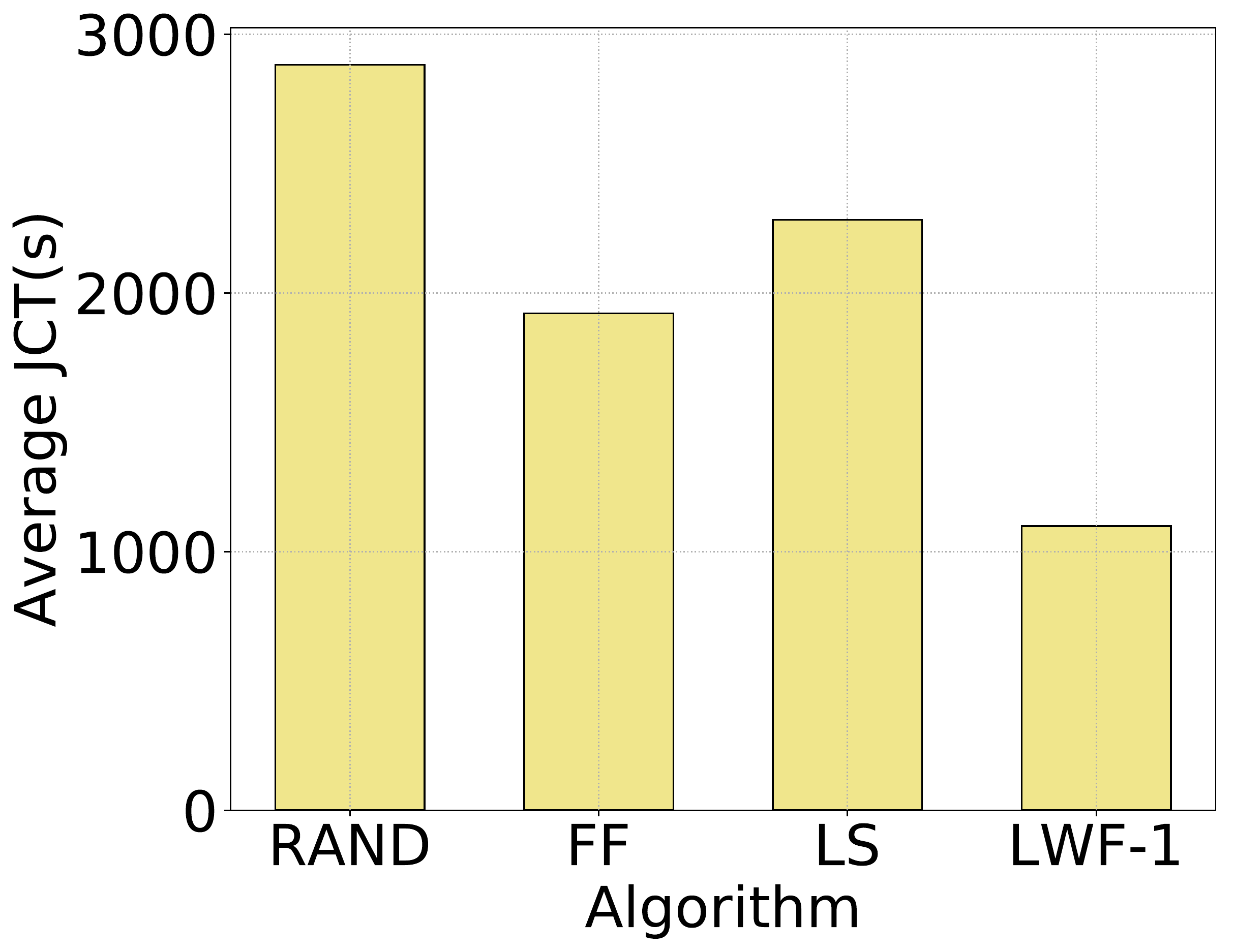}\label{fig:pg_c}
	}
	\caption{(a) The cumulative distribution of JCT of different placement algorithms. (b) The GPU utilization distribution of different placement algorithms. (c) The average JCT of different placement algorithms.}
	\vspace{-0.8 em}
	\label{fig:pg_results}
\end{figure*}

\textbf{The effectiveness of LWF-$\kappa$.} To fairly compare four placement algorithms, we apply the same Ada-SRSF scheduling algorithm for them. Fig. \ref{fig:pg_a} demonstrates the JCT distributions of four placement algorithms. Over 95\% of JCTs of LWF-$1$ are below 5000 seconds, which significantly outperforms the other three algorithms. RAND performs the worst because it easily selects the GPUs that have totally different workloads for a job and causes critically unbalanced GPU usage. Furthermore, FF surprisingly surpasses LS, although FF does not select those GPU with least workloads. The reason is that LS sometimes selects GPUs across too many servers, which introduces heavy communication overhead, while FF selects the GPUs in order following the first-fit strategy, which likely consolidates the GPUs together for a job. Fig. \ref{fig:pg_b} and Fig. \ref{fig:pg_c} respectively show the GPU utilization distributions and the average JCT of four algorithms. Our proposed LWF-$1$ also achieves the best results in terms of high GPU utilization and low average JCT. 
\begin{figure*}[!ht]
	\centering
	\subfigure[]
	{
	\includegraphics[width=0.29\linewidth]{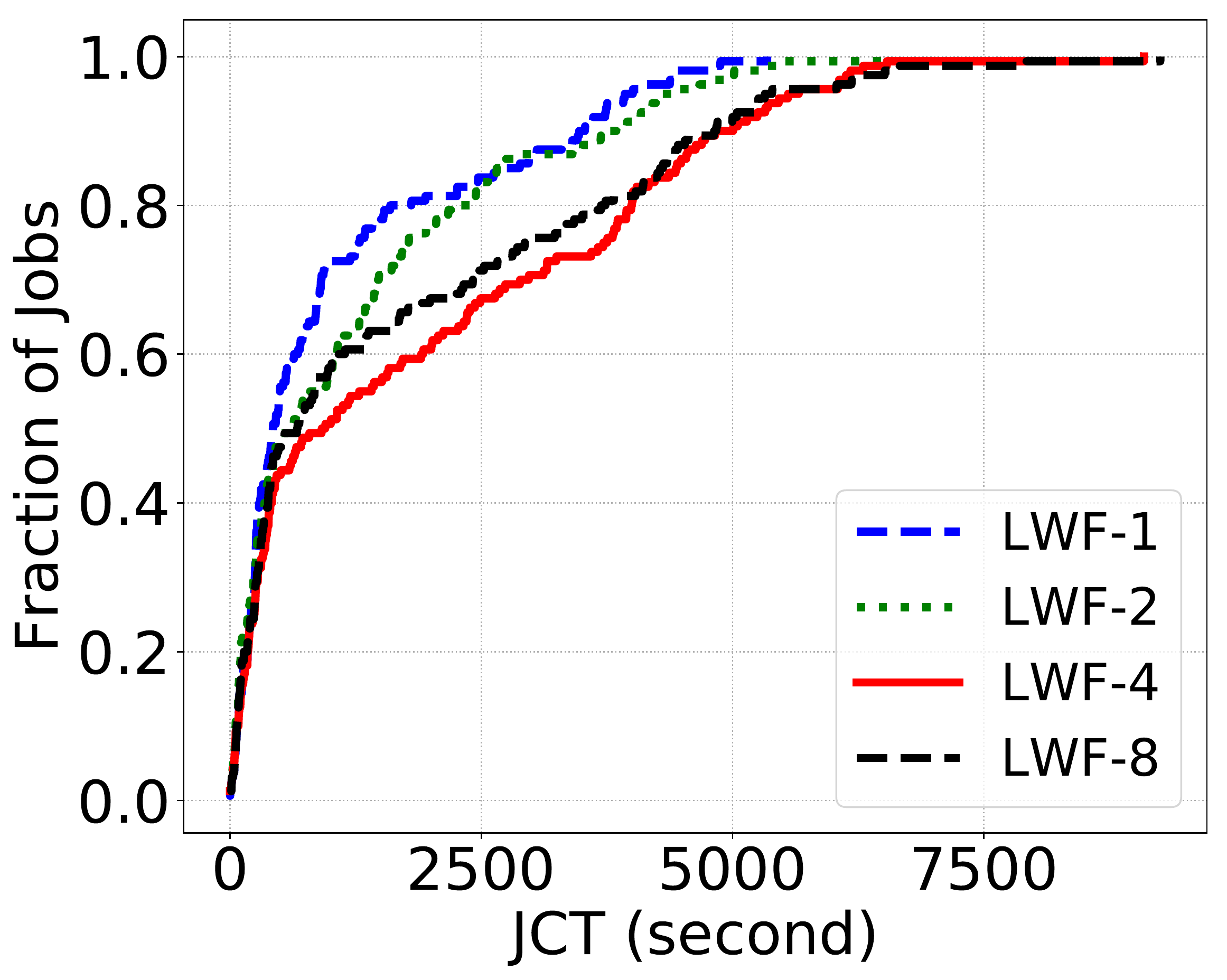}\label{fig:lwf_a}
	}
	\subfigure[]
	{
	\includegraphics[width=0.285\linewidth]{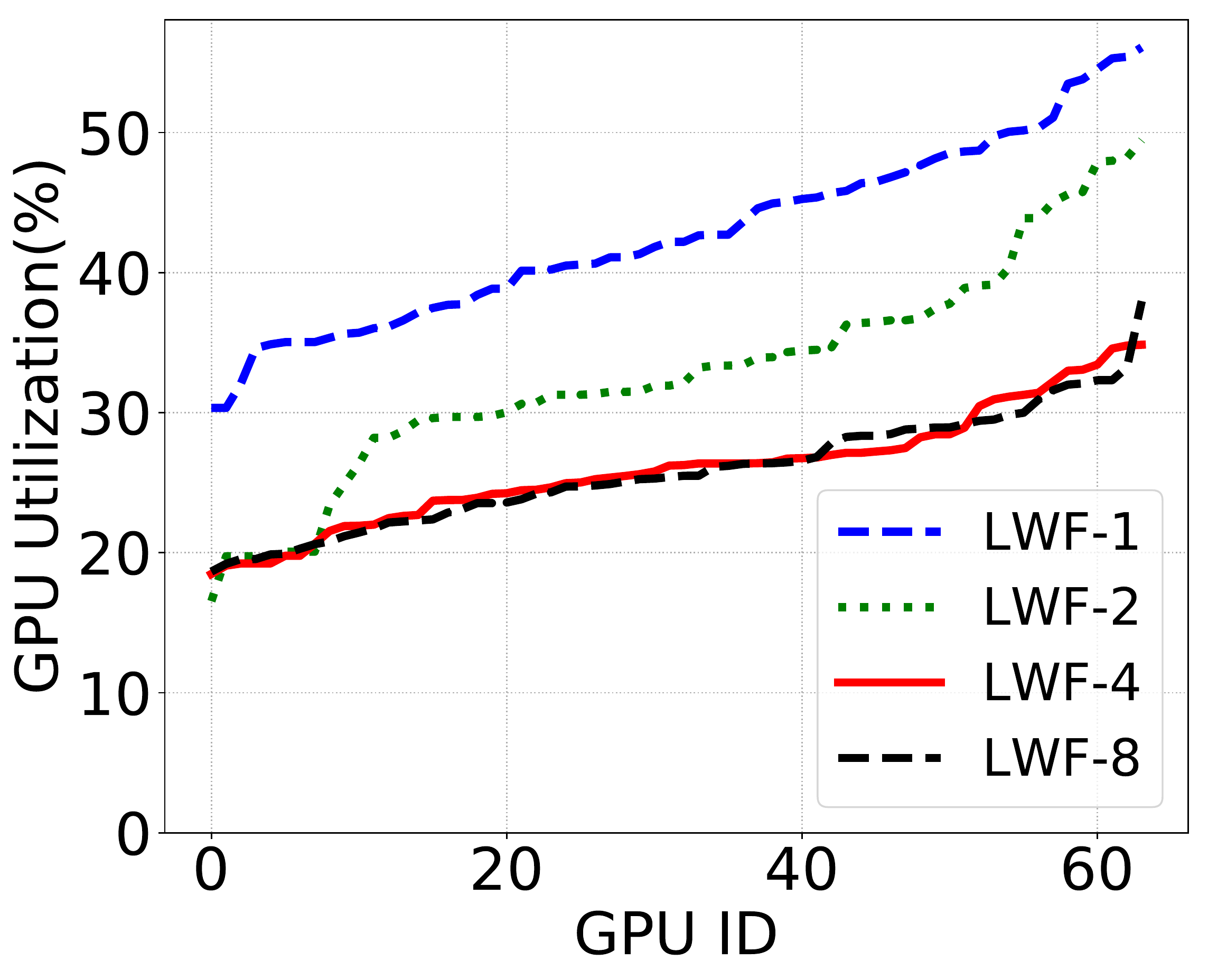}\label{fig:lwf_b}
	}
	\subfigure[]
	{
	\includegraphics[width=0.30\linewidth]{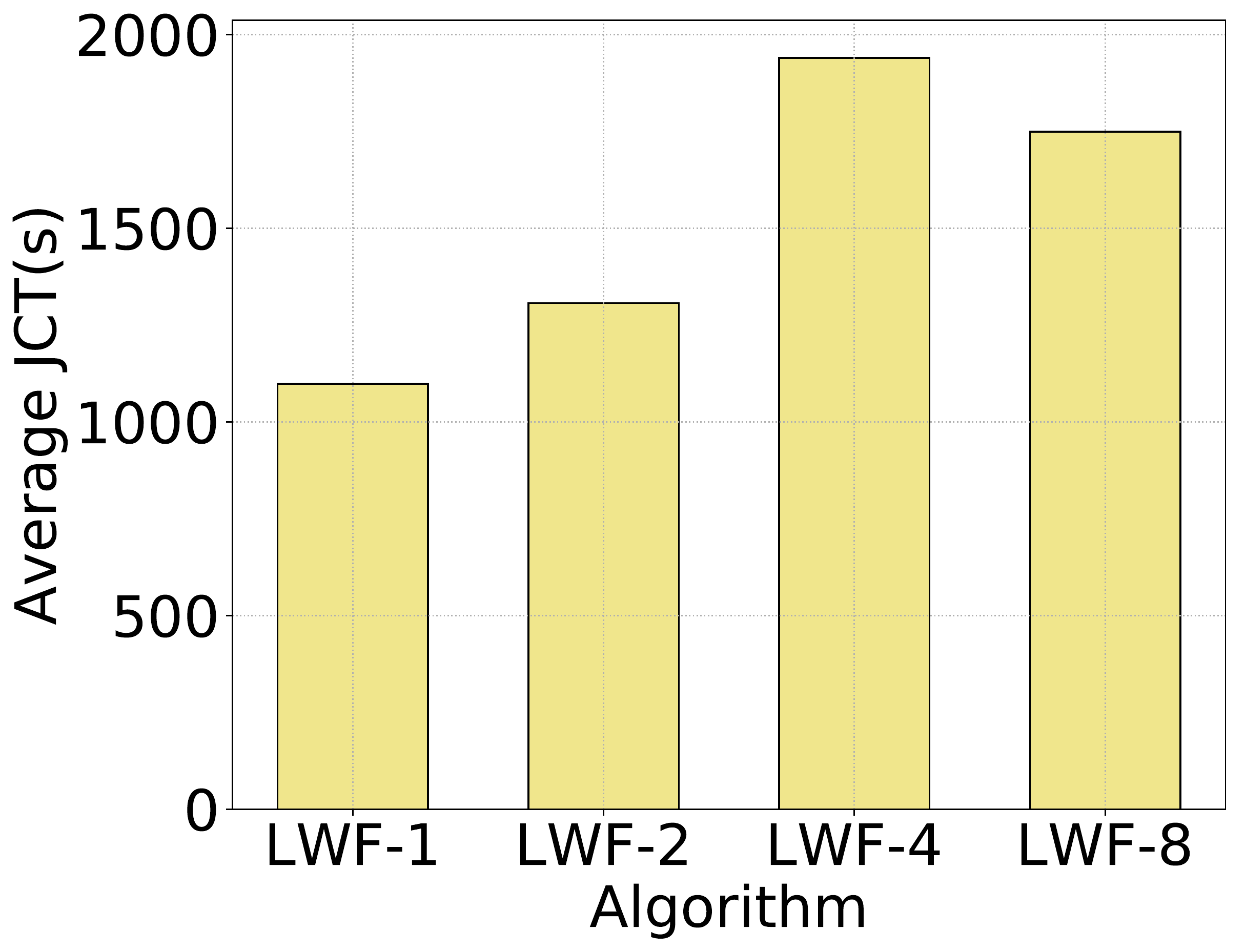}\label{fig:lwf_c}
	}
	\caption{(a) The cumulative distribution of JCT of different $\kappa$. (b) The GPU utilization distribution of different $\kappa$. (c) The average JCT of different $\kappa$.}
    \vspace{-0.8 em}
	\label{fig:lwf_results}
\end{figure*}

We also explore the impact of $\kappa$ to the LWF placement algorithm. Fig. \ref{fig:lwf_a}, \ref{fig:lwf_b} and \ref{fig:lwf_c} respectively show the JCT distributions, GPU utilization distributions and the average JCTs of different $\kappa$. It is observed that $\kappa=1$ generally gives the best results. As mentioned before, the threshold $\kappa$ is used to achieve two goals, GPU resource utilization balance and communication avoidance. Our results imply that for the 1-worker DDL job, it is better to choose the GPU with least workload globally, while for the others, choosing the GPUs server by server is more suitable.

Table \ref{tab:placement_results} concludes the experimental results of different placement solutions with Ada-SRSF. LWF-1 achieves 2.19$\times$, 1.59$\times$, 1.7$\times$ improvements in terms of average GPU utilization w.r.t to RAND, FF and LS. Besides, LWF-1 saves the average JCT by 61.9\%, 42.8\% and 51.9\% w.r.t to RAND, FF and LS. 
\begin{table}[!ht]
	\centering
	\caption{The experimental results of different placement solutions with Ada-SRSF.}
	\label{tab:placement_results}
		\begin{tabular}{|c|c|c|c|c|} \hline
			Method & Average & Average & Median & 95th \\ 
			       & GPU Util. & JCT(s) & JCT(s) & JCT(s) \\ \hline\hline
			RAND & 19.52\%	& 2881.6 & 680.3 & 8910.9 \\ \hline
			FF  & 26.76\%	& 1921.1 & 766.1 & 6224.6 \\ \hline
			LS  & 25.14\%	& 2282.41 & 749.9 & 7224.9 \\ \hline
			LWF-1  & \textbf{42.78\%} & \textbf{1098.57} & \textbf{487.2} & \textbf{4024.0} \\ \hline
		\end{tabular}
		\vspace{-0.8 em}
\end{table}

\begin{figure*}[!ht]
	\centering
	\subfigure[]
	{
	\includegraphics[width=0.305\linewidth]{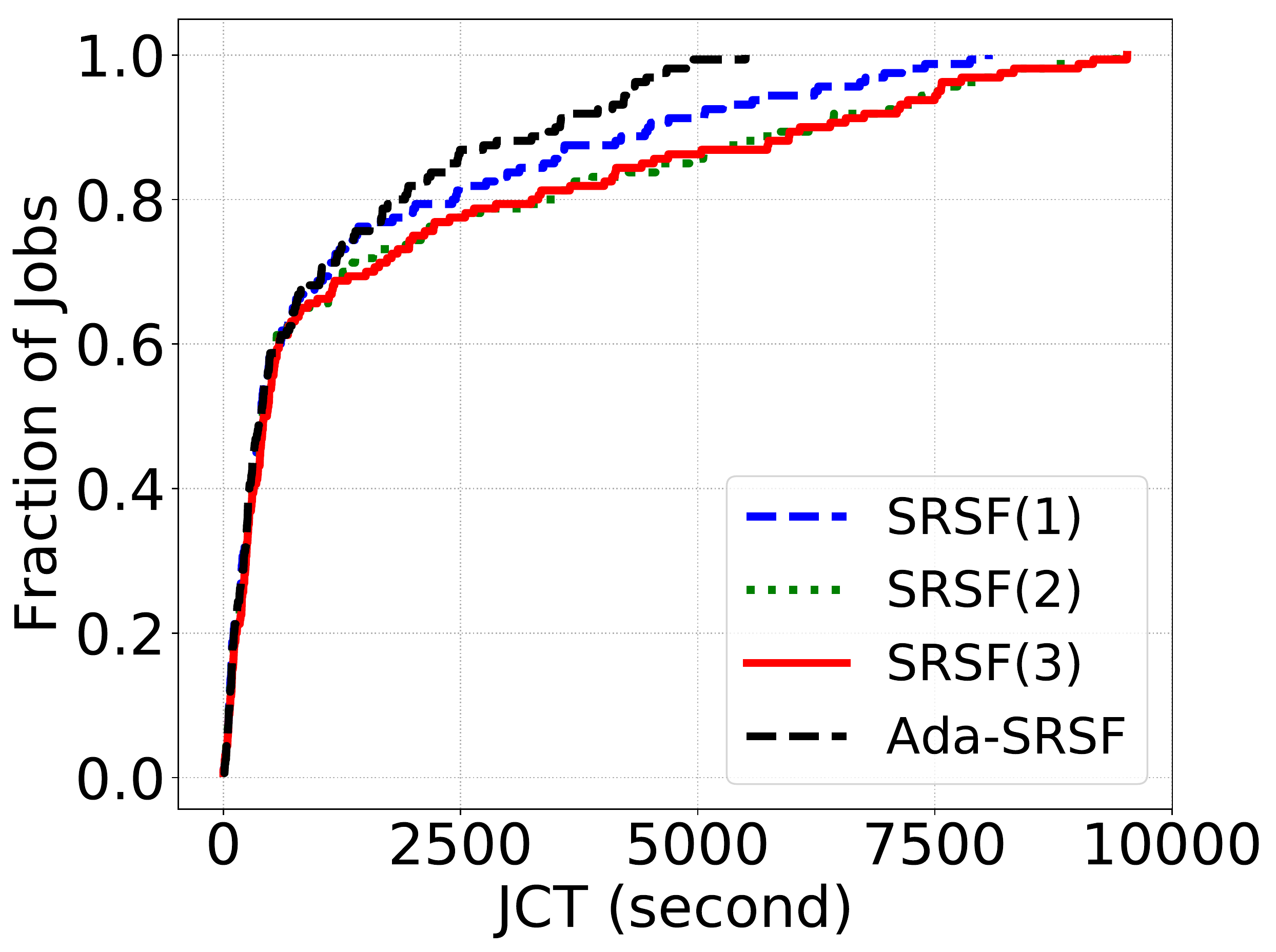}\label{fig:ada_a}
	}
	\subfigure[]
	{
	\includegraphics[width=0.28\linewidth]{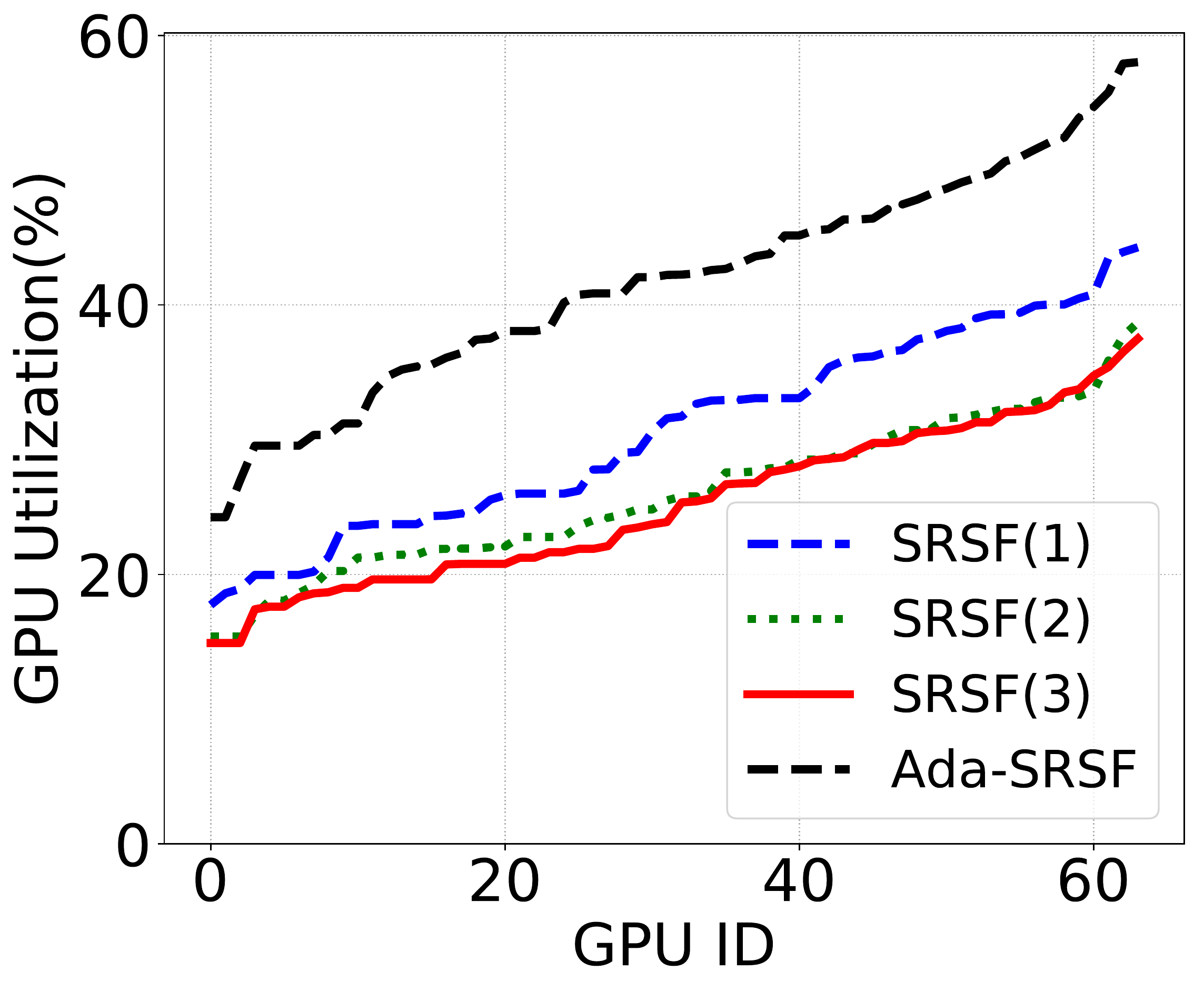}\label{fig:ada_b}
	}
	\subfigure[]
	{
	\includegraphics[width=0.30\linewidth]{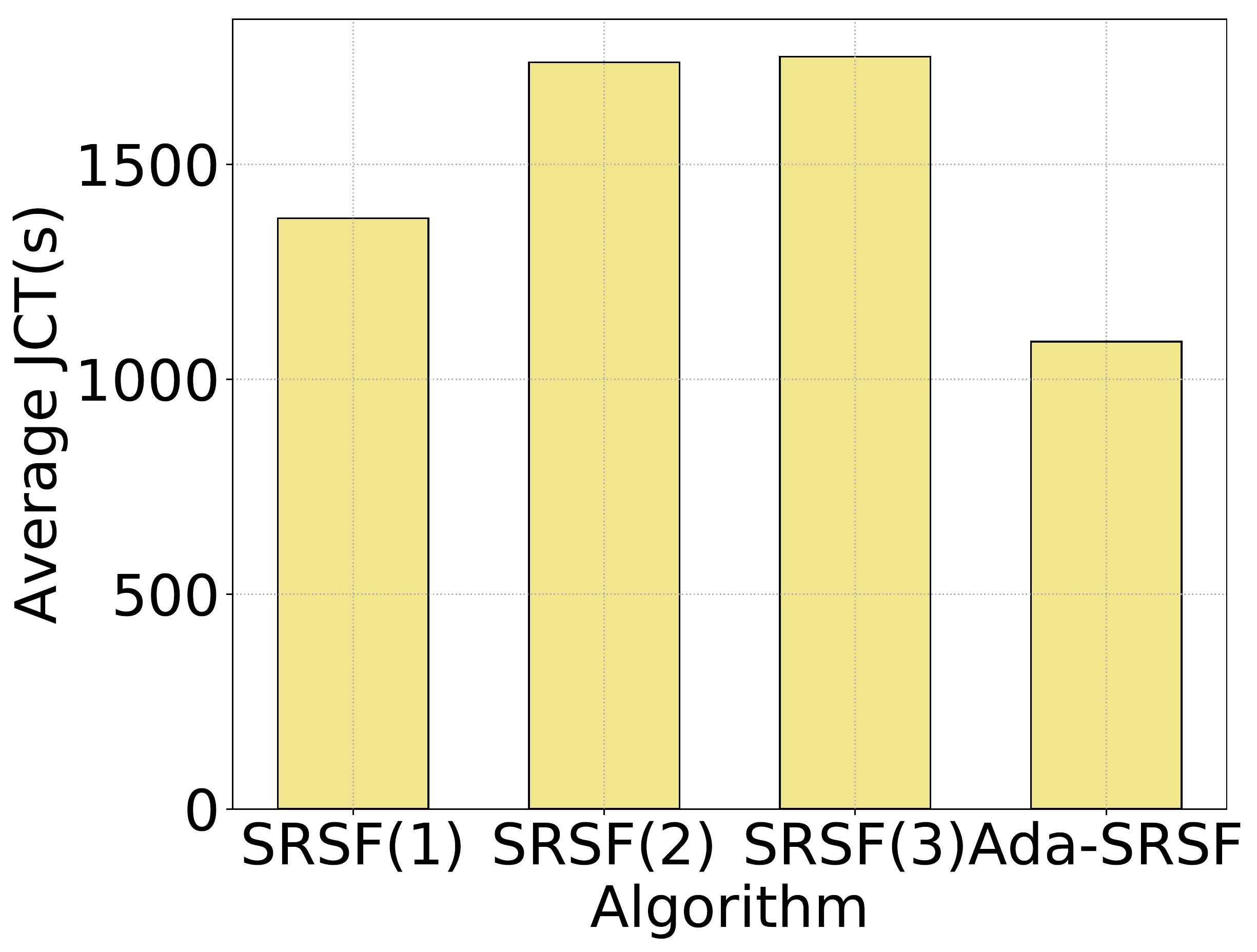}\label{fig:ada_c}
	}
	\caption{(a) The cumulative distribution of JCT of different communication scheduling algorithms. (b) The GPU utilization distribution of different communication scheduling algorithms. (c) The average JCT of different communication scheduling algorithms.}
	\label{fig:ada_results}
	\vspace{-0.8 em}
\end{figure*}

\textbf{The effectiveness of Ada-SRSF.} Fig. \ref{fig:ada_a} and \ref{fig:ada_b} respectively demonstrate the JCT distribution and the GPU utilization distribution of different communication scheduling algorithms. First, we observe that avoiding all the communication contention (SRSF(1)) is better than accepting them no matter the maximum number of allowing communication tasks is (SRSF(2) and SRSF(3)). This phenomenon suggests that simply adopting SRSF(1) generally achieves good performance. However, there still exists a big gap between SRSF(1) and Ada-SRSF. Notice that the GPU utilization of Ada-SRSF is much higher than that of SRSF(1). Ada-SRSF helps start some communication tasks earlier once it determines the advantages of communication contention, and then continues the GPU computation tasks. 

Table \ref{tab:schedule_results} concludes the experimental results of different communication scheduling algorithms with LWF-1. As for the average GPU utilization, Ada-SRSF achieves improvement of 39.6\% than SRSF(1). As for the average job completion time, Ada-SRSF saves 20.1\% time than SRSF(1). Notice that the 95th-percentile JCT of Ada-SRSF achieves 1.56$\times$ improvement w.r.t to SRSF(1), which explains the lower average JCT of Ada-SRSF. Although Ada-SRSF shows a considerable advantage among three scheduling algorithms, it needs more attention that the average GPU utilization is still less than 50\%. Since we do not exploit the possible overlapping of computation and communication of different layers in a fine-grained manner, the slow network can still become the system bottleneck and lower the GPU utilization. 
\begin{table}[!ht]
	\centering
	\caption{The experimental results of different communication scheduling solutions with LWF-1.}
	\label{tab:schedule_results}
		\begin{tabular}{|c|c|c|c|c|} \hline
			Method & Average & Average & Median & 95th \\ 
			       & GPU Util. & JCT(s) & JCT(s) & JCT(s) \\ \hline\hline
			SRSF(1) & 30.65\%	& 1374.84 & 459.0 & 6283.1 \\ \hline
			SRSF(2)  & 25.95\%	& 1734.74 & 477.0 & 7658.6 \\ \hline
			SRSF(3)  & 25.14\%	& 1750.9 & 520.7 & 7578.7 \\ \hline
			Ada-SRSF  & \textbf{42.78\%}	& \textbf{1098.57}  & \textbf{487.2} & \textbf{4024.0} \\ \hline
		\end{tabular}
	\vspace{-0.4 em}
\end{table}
\subsection{Discussion}
The simulation experiments based on the real measured hardware data reveal the great potentials of LWF-$\kappa$ and Ada-SRSF in the online DDL job scheduling problem. Theoretically, our solution can be applied to the real world applications. However, the existing popular deep learning software, such as PyTorch and Tensorflow\footnote{https://www.tensorflow.org}, and job management systems, such as Kubernetes\footnote{https://www.docker.com/topics/kubernetes}, do not support flexible task-level scheduling for DDL training. It is interesting to build a DDL job management system based on DAG formulation that allows task-level scheduling, which we leave as future work. 
\section{Related Work}\label{sec:relatedwork}

\textbf{Efficient DDL training system:} There have been efforts devoted to reducing the communication overhead for DDL training. %Those studies can be roughly categorized into two groups, pipeline overlap and workload reduction. Pipeline overlap leverages the parallelism of computation and communication during DDL training. 
Most of them achieve this target by exploiting the parallelism of computation and communication during DDL training and applying some well-designed communication strategies. Several open-sourced distributed training systems \cite{sdl2017,horovod,jia2018highly} exploit tensor fusion which merges those small sizes of gradients before aggregation among the trainers to reduce the communication overhead. The parameter server (PS) method \cite{ps2014} is one of the most representatives to adopt parallelism between computation and communication, but it easily suffers from the communication traffic jam of PS. In Poseidon proposed by Zhang et al. \cite{poseidon2017}, the wait-free backpropagation algorithm is invented to alleviate the communication pressure by overlapping communication and computation. Shi et al. \cite{mgwfbp2019,shi2020communication} propose an adaptive gradient fusion algorithm to further improve the wait-free backpropagation, which merges some small communicated data to reduce the overall communication time. Minsik et al. \cite{blueconnect2019} proposes BlueConnect, which decomposes a single all-reduce operation into a large number of parallelizable sub-operations to exploit the trade-off between latency and bandwidth, and adapt to different network environments. The communication overhead is a long-term obstacle for DDL training since the modern computing processors evolve much faster than the network device. Furthermore, while multiple DDL jobs are running on the cluster, it is more challenging to balance computation and communication, resolve resource contention and improve system utilization. 

%Workload reduction directly decreases the communicated data size, usually by gradient quantization and compression with maintaining model convergence, to lower the communication time. , topK and gtopK

\textbf{DDL training job scheduling system:} Recent years yield numerous studies about efficient DDL job schedulers and resource managers for a high-performance CPU/GPU-based clusters. They devote great efforts to promoting system utilization, minimizing the job completion time and saving the training budgets. Shi et al. \cite{shi2018dag} propose a DAG model for DDL training tasks for scheduling. Xiao. et al. \cite{gandiva2018} propose Gandiva, a resource manager for GPU clusters, which leverages the performance predictability of DDL training jobs to achieve better GPU time-slicing and job placement. Optimus \cite{optimus2018} is a similar resource scheduler to Gandiva but develops a DDL performance model given certain resources, and then accordingly adjusts resource allocation and job placement to achieve the lowest job completion time. Cynthia \cite{cynthia2019} improves the DDL performance model by integrating the effects of system heterogeneity and covering both synchronous and asynchronous training. It also develops a resource provisioning strategy that not only optimizes the system performance, but also saves the training budget. However, these strategies rely on time prediction methodology which makes over-simplified assumptions of the training convergence curves. They are usually not applicable to practical production systems. Tiresias \cite{tiresias2019} assumes the unpredictability of the DDL job performance and develops a GPU cluster manager for DDL training only based on the overall distribution of the historical job duration. It applies a two-dimension scheduling algorithm, which considers both job attained service time and its occupied GPUs, to decide the job priority. Mei et al. \cite{xin2017scheduling} and Chau et al. \cite{chau2017scheduling} proposed scheduling algorithms for CPU-GPU heterogeneous clusters, which aimed for maximizing the energy efficiency without severe performance sacrifice. However, none of them considers the extra overhead of potential communication resource competition. 
%When network matters \cite{network2019}, makespan of DAG jobs , 

Different from traditional heuristic solutions based on optimization techniques and experimental observations, some studies take advantage of the learning-based methods to solve the task scheduling and placement problem. Bao. et al. \cite{dlblb2019} presents Harmony, a deep learning-driven ML cluster scheduler that places training jobs in a manner that minimizes interference and maximizes performance, such as training completion time. Hu. \cite{spear2019} develops Spear, a new scheduling framework designed to minimize the makespan of complex jobs, while considering both task dependencies and heterogeneous resource demands at the same time. 

\section{Conclusion and Future Work}\label{sec:conclusion}
In this paper, we first claimed that the limited network resources usually restrict the scalability and efficiency of multiple DDL training jobs on the cloud clusters. Even the common communication contention may drastically bring down the resource utilization and finally increase the job completion. Different from the previous studies, we establish a new DDL scheduling framework which organizes the DDL jobs as DAGs for scheduling flexibility. Our new problem formulation takes into account the effects of limited network resources and potential communication contention. We derive an efficient and effective job placement and scheduling algorithms, LWF-$\kappa$ and Ada-SRSF, to give an elegant solution to the problem. We conducted experiments with simulation on a 16-server cluster connected with 10GbE links, and each node is equipped with four Nvidia GPUs. Experimental results showed that our proposed LWF-$\kappa$ placement algorithm outperforms existing first-fit and list-scheduling heuristic algorithms. Ada-SRSF shows great improvement of time efficiency and resource utilization over the solutions of avoiding all the contention and accepting all of them blindly. 

There are four directions for the future work. Firstly, as for the placement, we set a threshold $\kappa$ to balance the resource utilization and communication overhead, which depends on the empirical study and lacks theoretical guarantee. A further theoretical analysis of the placement problem should be derived and helps produce a potentially better solution. Secondly, as for the scheduling, we only discuss the case that a maximum of two-way communication contention is allowed. Naturally it is worthy to explore efficient solutions to the cases of $k$-way communication contention when $k$ is larger than two. Thirdly, it is interesting to introduce layer-wise techniques into the job scheduling framework and derive efficient scheduling algorithms by overlapping the computation and communication of different layers. Finally, maximizing the energy efficiency of deep learning systems also deserves exploration.

\section*{Acknowledgements}
This research was supported by Hong Kong RGC GRF grant HKBU 12200418. We thank the anonymous reviewers for their constructive comments and suggestions. We would also like to thank NVIDIA AI Technology Centre (NVAITC) for providing the GPU clusters for some experiments.

\bibliographystyle{IEEEtran}
%\bibliography{dl_scheduling}
\bibliography{Communication-Contention-Aware-Scheduling-of-Multiple-Deep-Learning-Training-Jobs.bbl}

\end{document}